\documentclass[12pt]{article}
\usepackage{epsfig,rotating,setspace,latexsym,amsmath,epsf,amssymb,amsfonts,bm,bbm,theorem,cite,caption,subcaption,enumerate,longtable,accents,theoremref,mathtools,mathrsfs}
\usepackage{algorithm,algorithmic,graphicx,epsf,authblk,epstopdf,url,color,multirow}
\usepackage{soul}
\usepackage{tabularx}
\usepackage{booktabs}

\newtheorem{theorem}{Theorem}

\newtheorem{lemma}{Lemma}
\newenvironment{Proof}[1]{\medskip\par\noindent{\bf Proof:\,}\,#1}{{\mbox{\,$\blacksquare$}\par}}
\newcounter{example}

\DeclarePairedDelimiter{\ceil}{\lceil}{\rceil}
\allowdisplaybreaks

\begin{document}

\title{Group Testing with a Graph Infection Spread Model\thanks{This work is presented in part at IEEE ISIT, July 2021.}}
\author{Batuhan Arasli \qquad Sennur Ulukus\\
	\normalsize Department of Electrical and Computer Engineering\\
	\normalsize University of Maryland, College Park, MD 20742 \\
	\normalsize {\it barasli@umd.edu} \qquad {\it ulukus@umd.edu}}

\date{}
\maketitle

\vspace*{-1.0cm}

\begin{abstract}
We propose a novel infection spread model based on a random connection graph which represents connections between $n$ individuals. Infection spreads via connections between individuals and this results in a probabilistic cluster formation structure as well as a non-i.i.d.~(correlated) infection status for individuals. We propose a class of \emph{two-step sampled group testing algorithms} where we exploit the known probabilistic infection spread model. We investigate the metrics associated with two-step sampled group testing algorithms. To demonstrate our results, for analytically tractable \emph{exponentially split cluster formation trees}, we calculate the required number of tests and the expected number of false classifications in terms of the system parameters, and identify the trade-off between them. For such exponentially split cluster formation trees, for zero-error construction, we prove that the required number of tests is $O(\log_2n)$. Thus, for such cluster formation trees, our algorithm outperforms any zero-error non-adaptive group test, binary splitting algorithm, and Hwang's generalized binary splitting algorithm. Our results imply that, by exploiting probabilistic information on the connections of individuals, group testing can be used to reduce the number of required tests significantly even when infection rate is high, contrasting the prevalent belief that group testing is useful only when infection rate is low.
\end{abstract}

\section{Introduction}
The group testing problem, introduced by Dorfman in \cite{dorfman1943}, is the problem of identifying the infection status of a set of individuals by performing fewer tests than individually testing everyone. The key idea of group testing is to mix test samples of the individuals and test the mixed sample. A negative test result implies that everyone within that group is negative, thereby identifying infection status of an entire group with a single test. A positive test result implies that there is at least one positive individual in that group, in which case, Dorfman's original algorithm goes into a second phase of testing everyone individually. 

Since Dorfman's seminal work, various families of algorithms have been studied, such as, adaptive algorithms, where one designs test pools in the $(i+1)$st step by using information from the test results in the first $i$ steps, and non-adaptive algorithms, where every test pool is predetermined and run in parallel. In addition, various forms of infection spread models have been considered as well, such as, the independent and identically distributed (i.i.d.) model where each person is infected independent of others with probability $p$, and the combinatorial model where $k$ out of $n$ people are infected uniformly distributed on the sample space of $\binom{n}{k}$ elements. Under these various system models and family of algorithms, the group testing problem has been widely studied. For instance, \cite{combinatorial_gt} gives a detailed study of combinatorial group testing and zero-error group testing, \cite{bornagain_mac} relates the group testing problem to a channel coding problem, and \cite{atia_saligrama_first, wadayama_nonadaptive, wang_combquant, wu_partition, shangguan_newbounds, scarlett_noisynonadaptive, scarlett_noisy_separate, scarlett_noisyadaptivebounds, mazumdar_nonadaptive, kealy_capacity, johnson_nearconstant, inan_optimalityks, karimi_irregularsparsegraph, gebhard2019quantitative, falahatgar, coja_threshold, nonadaptive_bounds, cai_noisy, bondorf2020sublineartime, aldridge_individual, impossibilityresults, heidarzadeh2020twostage} advance the group testing literature in various directions. The advantage of group testing is known to diminish when the disease is not rare \cite{RUSZINKO, sharper, scarlettbook}.

Early works mainly consider two infection models: combinatorial model where prior to designing the algorithm the exact number of infections is assumed to be known, and probabilistic model where each individual is assumed to be infected with probability $p$ identically and independently. Although there is no general result for arbitrary infection probabilities and arbitrary correlations, the following papers have considered advanced probabilistic models: \cite{prior_stats} considers a probabilistic model with independent but non-identically distributed infection probabilities.  \cite{correlated_bio} considers a correlated infection distribution  under very specific assumptions. \cite{lincorrelated} considers a system where individuals are modelled as a community with positive correlations between them for specific setups, such as individuals at contiguous positions in a line. \cite{community_aware} considers a model where individuals belong to disjoint communities, and the system parameters are the number of infected families and the probability that a family is infected. Authors show that leveraging the community information improves the testing performance by reducing the number of tests required, from the scale of number of infections to the scale of number of infected families for both probabilistic and combinatorial setups. In the subsequent work \cite{diggavioverlap}, authors consider overlapping communities. In \cite{community_structure}, authors focus on community structured system model, where the underlying network model is drawn from the stochastic block model. Over a fixed community structure, initial infections are introduced i.i.d.~to the system, then infection spread within and between communities are realized, with infections spreading within community with higher fixed probability than between communities. Authors propose an adaptive algorithm and compare its performance with the binary splitting algorithm that does not leverage the community information. Our goal in this paper is to consider a realistic graph-based infection spread model, and exploit the knowledge of the infection spread model to design efficient group testing algorithms.

To that end, first, we propose a novel infection spread model, where individuals are connected via a random connection graph, whose connection probabilities are known\footnote{For instance, location data obtained from cell phones can be used to estimate connection probabilities.}. A realization of the random connection graph results in different connected components, i.e., clusters, and partitions the set of all individuals. The infection starts with a patient zero who is uniformly randomly chosen among $n$ individuals. Then, any individual who is connected to at least one infected individual is also infected. For this system model, we propose a novel family of algorithms which we coin \emph{two-step sampled group testing algorithms}. The algorithm consists of a sampling step, where a set of individuals are chosen to be tested, and a zero-error non-adaptive test step, where selected individuals are tested according to a zero-error non-adaptive group test matrix. In order to select individuals to test in the first step, one of the possible cluster formations that can be formed in the random connection graph, is selected. Then, according to the selected cluster formation, we select exactly one individual from every cluster. After identifying the infection status of the selected individuals with zero-error, we assign the same infection status to the other individuals in the same cluster with identified individuals. Note that, the actual cluster formation is not known prior to the test design, and because of that, selected cluster formation can be different from the actual cluster formation. Thus, this process is not necessarily a zero-error group testing procedure.

Our main contributions consist of proposing a novel infection spread model with random connection graph, proposing a two-step sampled group testing algorithm which is based on novel $\mathcal{F}$-separable zero-error non-adaptive test matrices, characterizing the optimal design of two-step sampled group testing algorithms, and presenting explicit results on analytically tractable \emph{exponentially split cluster formation trees}. For the considered two-step sampled group testing algorithms, we identify the optimal sampling function selection, calculate the required number of tests and the expected number of false classifications in terms of the system parameters, and identify the trade-off between them. Our $\mathcal{F}$-separable zero-error non-adaptive test matrix construction is based on taking advantage of the known probability distribution of cluster formations. In order to present an analytically tractable case study for our proposed two-step sampled group testing algorithm, we consider exponentially split cluster formation trees as a special case, in which we explicitly calculate the required number of tests and the expected number of false classifications. For zero-error construction, we prove that the required number of tests is less than $4(\log_2n+1)/3$ and is of $O(\log_2n)$, when there are at most $n$ equal-sized clusters in the system, each having $\delta$ individuals. For the sake of fairness, in our comparisons, we take $\delta$ to be 1, ignoring further reductions of the number of tests due to $\delta$. We show that, even when we ignore the gain by cluster size $\delta$, our non-adaptive algorithm, in the zero-error setting, outperforms any zero-error non-adaptive group test and Hwang's generalized binary splitting algorithm \cite{hwang_binary}, which is known to be the optimal zero-error adaptive group test \cite{scarlettbook}. Since the number of infections scale as $\frac{n}{\log_2n}\delta$ in exponentially split cluster formation trees with $n\delta$ individuals, our results show that, we can use group testing to reduce the required number of tests significantly in our system model even when the infection rate is high by using our two-step sampled group testing algorithm.

\section{System Model} \label{sec2}
We consider a group of $n$ individuals. The \emph{random infection vector} $U=(U_1,U_2,\dots ,U_n)$ represents the infection status of the individuals. Here $U_i$ is a Bernoulli random variable with parameter $p_i$. If individual $i$ is infected then $U_i = 1$, otherwise $U_i=0$. Random variables $U_i$ need not be independent. A \emph{patient zero random variable} $Z$ is uniformly distributed over the set of individuals, i.e., $Z=i$ with probability $p_Z(i)=\frac{1}{n}$ for $i=1,\ldots,n$. Patient zero is the first person to be infected. So far, the infection model is identical to the traditional combinatorial model with $k=1$ infected among $n$ individuals.

\begin{table} [h]
\caption{Nomenclature.}
\centering
\begin{tabularx} {\textwidth}{|c|>{\raggedright\arraybackslash}X|}
    \hline
    \multicolumn{2}{|c|}{\textbf{System}}\\
    \hline
    $n$ & number of individuals in the system \\
    $U$ & infection status vector of size $n$\\
    $Z$ & patient zero random variable\\
    $p_Z(i)$ &probability of individual i is the patient zero\\
    $\mathscr{C}$ & random connection graph\\
    $E_\mathscr{C}$ & edge set of $\mathscr{C}$\\
    $V_\mathscr{C}$ & vertex set of $\mathscr{C}$, also equal to $[n]$\\
    $\bm{C}$ & random connection matrix\\
    $F$ &cluster formation random variable\\
    $\mathcal{F}$ &set of all possible cluster formations, i.e., $\{F_i\}$\\
    $p_F(F_i)$ &probability of true cluster formation is $F_i$\\
    $f$ &number of possible cluster formations, i.e., $|\mathcal{F}|$\\
    $\sigma_i$ &number of clusters in the cluster formation $F_i$\\
    $S_j^i$ & $j${th} cluster in $F_i$\\
    $\lambda_j$ & number of unique clusters in $\mathcal{F}$ at and above the level $F_j$\\
    $\lambda_{S_i^j}$ & number of unique ancestor nodes of $S_i^j$ in $\mathcal{F}$\\
    $\delta$ & size of the bottom level clusters in an exponentially split $\mathcal{F}$ \\
    \hline
    \multicolumn{2}{|c|}{\textbf{Algorithm}}\\
    \hline
    $F_m$   & sampling cluster formation chosen from $\mathcal{F}$\\
    $M$ &sampling function that selects individuals to be tested\\
    $U^{(M)}$ &infection status vector of the selected individuals by $M$ \\
    $S^\alpha(M_i)$ & the cluster in $F_\alpha$ that contains $i${th} selected individual by $M$\\
    $K_M$ & set of infections among the selected individuals by $M$\\
    $\mathcal{P}(K_M)$ & set of all possible infected sets that $K_M$ can be\\
    $T$ & number of tests to be performed\\
    $\bm{X}$ & $T \times \sigma _m$ test matrix\\
    $\bm{X}^{(i)}$ &$i${th} column of $\bm{X}$ \\
    $y$ &test result vector of size $T$\\
    $\hat{U}$ &estimated infection status of $n$ individuals after test results\\
    $E_{f,\alpha}$ &expected number of false classifications given $F=F_\alpha$\\
    $E_f$ &expected number of false classifications\\
    \hline
\end{tabularx}
\end{table}

Next, we define a \emph{random connection graph} $\mathscr{C}$ which is a random graph where vertices represent the individuals and edges represent the connections between the individuals. Let $p_{\mathscr{C}}$ denote the probability distribution of the random graph $\mathscr{C}$ over the support set of all possible edge realizations. For the special class of random connection graphs where the edges are realized independently, we fully characterize the statistics of the random connection graph by the random connection matrix $\bm{C}$, which is a symmetric $n \times n$ matrix where the $(i,j)$th entry $C_{ij}$ is the probability that there is an edge between vertices $i$ and $j$ for $i\neq j$, and $C_{ij}=0$ for $i=j$ by definition.
	
A random connection graph $\mathscr{C}$ is an undirected random graph with vertex set $V_\mathscr{C} = [n]$, with each vertex representing a unique individual, and a random edge set $E_\mathscr{C} = \{e_{ij}\}$ which represents connections between individuals, that satisfies the following: 1) If $e_{ij} \in E_\mathscr{C}$, then there is an edge between vertices $i$ and $j$; 2) For an arbitrary edge set $E_\mathscr{C}^*$, probability of $E_\mathscr{C}=E_\mathscr{C}^*$ is equal to $p_\mathscr{C}(E_\mathscr{C}^*,V_\mathscr{C})$. In the case when all $\mathbbm{1}_{\{e_{ij} \in E_\mathscr{C}\}}$ are independent, where $\mathbbm{1}_{A}$ denotes the indicator function of the event $A$, the random connection matrix $\bm{C}$ fully characterizes the statistics of edge realizations. There is a path between vertices $i$ and $j$ if there exists a set of vertices $\{i_1, i_2, \dots i_k \}$ in $[n]$ such that $\{e_{ii_1}, e_{i_1i_2}, e_{i_2i_3}, \dots e_{i_kj} \} \subset E_\mathscr{C}$, i.e., two vertices are connected if there exists a path between them.
	
In our system model, if there is a path in $\mathscr{C}$ between two individuals, then their infection status are equal. In other words, the infection spreads from patient zero $Z$ to everyone that is connected to patient zero. Thus, $U_k = U_l$ if there exists a path between $k$ and $l$ in $\mathscr{C}$. Here, we note that, a realization of the random graph $\mathscr{C}$ consists of clusters of individuals, where a cluster is a subset of vertices in $\mathscr{C}$ such that all elements in a cluster are connected with each other and none of them is connected to any vertex that is not in the cluster. More rigorously, a subset $S = \{i_1, i_2, \dots i_k\}$ of $V_\mathscr{C}$ is a cluster if, $i_l$ and $i_m$ are connected for all $i_l \neq i_m \in S$, but $i_a$ and $i_b$ are not connected for any $i_a \in S$ and all $i_b \in V_\mathscr{C} \backslash S$.
	
Note that the set of all clusters in a realization of the random graph $\mathscr{C}$ is a partition of $[n]$. In a random connection graph structure, formation of clusters in $\mathscr{C}$ along with patient zero $Z$ determine the status of the infection vector. Therefore, instead of focusing on the specific structure of the graph $\mathscr{C}$, we focus on the cluster formations in $\mathscr{C}$. For a given $p_\mathscr{C}$, we can calculate the probabilities of possible cluster formations in $\mathscr{C}$.

\begin{figure} [t]
	\centering
	\begin{subfigure}[b]{0.475\textwidth}
		\centering
		\epsfig{file=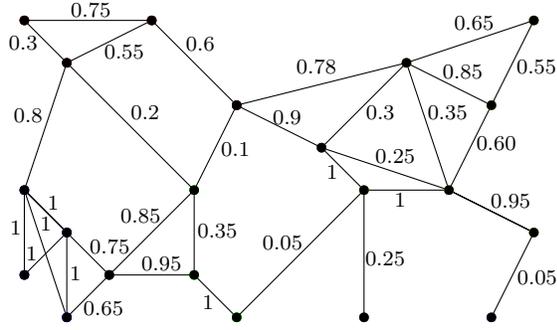}
		\caption[]%
		{{\small Probabilities of the edges.}}    
	\end{subfigure}
	\hfill
	\begin{subfigure}[b]{0.475\textwidth}  
		\centering 
		\epsfig{file=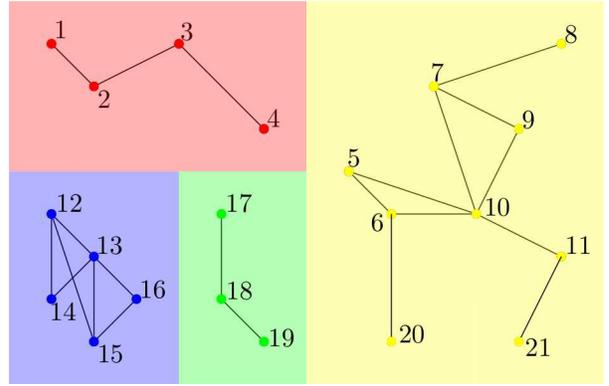}
		\caption[]%
		{{\small In this realization of $\mathscr{C}$, there are 4 clusters.}}
	\end{subfigure}
	\vskip\baselineskip
	\begin{subfigure}[b]{0.475\textwidth}   
		\centering 
		\epsfig{file=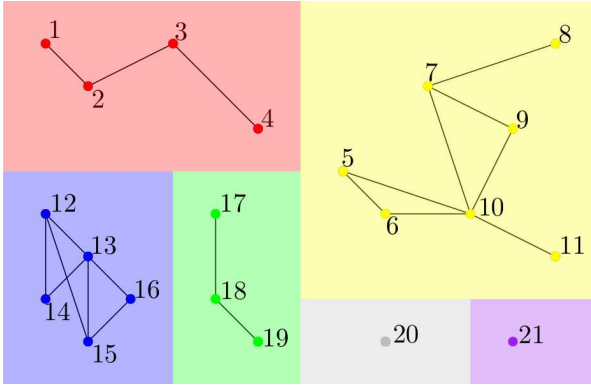}
		\caption[]%
		{{\small In this realization of $\mathscr{C}$, there are 6 clusters.}}
	\end{subfigure}
	\hfill
	\begin{subfigure}[b]{0.475\textwidth}   
		\centering 
		\epsfig{file=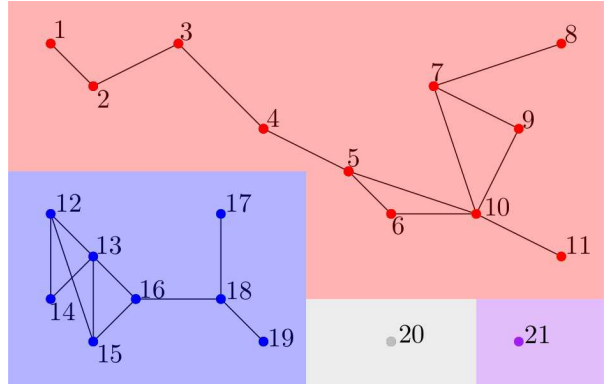}
		\caption[]%
		{{\small In this realization of $\mathscr{C}$, there are 4 clusters.}}
	\end{subfigure}
	\caption{Random connection graph $\mathscr{C}$ and three possible realizations and cluster formations. We show each cluster with a different color.} 
	\label{Crealizationsfig}
\end{figure}

To solidify ideas, we give an example in Figure~\ref{Crealizationsfig}. For a random connection graph where the edges are realized independently, we give probabilities of the existence of edges (zero probabilities are not shown) in Figure~\ref{Crealizationsfig}(a) and three different realizations of a random connection graph $\mathscr{C}$, where all three realizations result in different cluster formations in Figure~\ref{Crealizationsfig}(b)-(d). In Figure~\ref{Crealizationsfig}, we consider a random connection graph $\mathscr{C}$ that has $n=21$ vertices, which represent the individuals in our group testing model. Since in this example we assume that the edges are realized independently, every edge between vertices $i$ and $j$ exists with probability $C_{ij}$, independently. As we defined, if there is a path between two vertices (i.e., they are in the same cluster), then we say that their infection status are the same. One way of interpreting this is, there is a patient zero $Z$, which is uniformly randomly chosen among $n$ individuals, and patient zero spreads the infection to everyone in its cluster. Therefore, working on the cluster formation structures, rather than the random connection graph itself, is equally informative for the sake of designing group tests. For instance, in the realization that we give in Figure~\ref{Crealizationsfig}(b), if the edge between vertices 5 and 10 did not exist, that would be a different realization for the random connection graph $\mathscr{C}$, however, the cluster formations would still be the same. As all infections are determined by the cluster formations and the realization of patient zero, cluster formations are a sufficient statistics. Before we rigorously argue this point, we first focus on constructing a basis for random cluster formations.

The \emph{random cluster formation} variable $F$ is distributed over $\mathcal{F}$ as $\mathbb{P}(F=F_i) = p_F(F_i)$, for all $F_i \in \mathcal{F}$, where $\mathcal{F}$ is a subset of the set of all partitions of the set $\{1, 2, \dots, n\}$. In our model, we know the set $\mathcal{F}$ (i.e., the set of cluster formations that can occur) and the probability distribution $p_F$, since we know $p_\mathscr{C}$. Let us denote $|\mathcal{F}|$ by $f$. For a cluster formation $F_i$, individuals that are in the same cluster have the same infection status. Let $|F_i| = \sigma_i$, i.e., there are $\sigma_i$ subsets in the partition $F_i$ of $\{1,2, \dots ,n\}$. Without loss of generality, for $i<j$, we have $\sigma_i \leq \sigma_j$, i.e., cluster formations in $\mathcal{F}$ are ordered in increasing sizes. Let $S_j^i$ be the $j$th subset of the partition $F_i$ where $i \in [f]$ and $j \in [\sigma_i]$. Then, for fixed $i$ and $j$, $U_k = U_l$ for all $k,l \in S_j^i$, for all $i \in [f]$ and $j \in [\sigma_i]$. 

\begin{figure} [t]
    \centering
    \epsfig{file=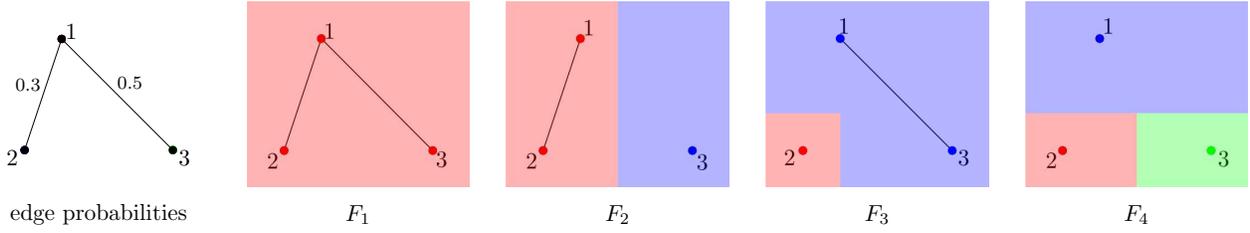,width=1\textwidth}
    \caption{Edge probabilities of $\mathscr{C}$ and elements of $\mathcal{F}$ in example $\bm{C}$ given in (\ref{example1}) with clusters shown in different colors.}
    \label{fig7}
    \vspace*{-0.4cm}
\end{figure}

To clarify the definitions, we give a simple running example which we will refer to throughout this section. Consider a population with $n=3$ individuals who are connected according to the random connection matrix $\bm{C}$ and assume that the edges are realized independently,  
\begin{align} \label{example1}
    \bm{C}= \begin{bmatrix}
            0   & 0.3 & 0.5\\
            0.3 & 0   & 0\\
            0.5 & 0   & 0
           \end{bmatrix}
\end{align}
By definition, the main diagonal of the random connection matrix is zero, since we define edges between distinct vertices only. In this example, $\mathcal{F}$ consists of 4 possible cluster formations, and thus, we have $f=|\mathcal{F}|=4$. The random cluster formation variable $F$ can take those 4 possible cluster formations with following probabilities,
\begin{align} \label{example1-cluster}
F = \begin{cases}
	F_1=\{\{1,2,3\}\}, \enskip &\textit{with probability } 0.15 \\
	F_2=\{\{1,2\},\{3\}\}, \enskip &\textit{with probability } 0.15 \\
	F_3=\{\{1,3\},\{2\}\}, \enskip &\textit{with probability } 0.35 \\
	F_4=\{\{1\},\{2\},\{3\}\}, \enskip &\textit{with probability } 0.35 \\
	\end{cases}
\end{align}
This example network and the corresponding cluster formations are shown in Figure~\ref{fig7}. Here, cluster formation $F_1$ occurs when the edge between vertices 1 and 2 and the edge between vertices 1 and 3 are realized; $F_2$ occurs when only the edge between vertices 1 and 2 is realized; and $F_3$ occurs when only the edge between vertices 1 and 3 is realized. Finally, $F_4$ occurs when none of the edges in $\mathscr{C}$ is realized. In this example, we have $\sigma_1 = |F_1| = 1$, $\sigma_2 = |F_2| = 2$, $\sigma_3 = |F_3| = 2$, and $\sigma_4 = |F_4| = 3$. Note that $\sigma_1\leq \sigma_2\leq \sigma_3\leq \sigma_4$ as assumed without loss of generality above. Each subset that forms the partition $F_i$ are denoted by $S_j^i$, for instance, $F_3$ consists of $S_1^3 = \{1,3\}$ and $S_2^3=\{2\}$. 

Next, we argue formally that cluster formations are a sufficient statistics, i.e., they represent equal amount of information as the realization of the random graph as far as the infection status of the individuals are concerned. When $Z$ and $F$ are realized, the infection status of $n$ individuals are also realized, i.e., $H(U|Z,F) = 0$. Then,
\begin{align}
	I(U;F) &= H(U)-H(U|F)\\
		   &= H(U)-\left(H(U,Z|F)-H(Z|U,F)\right)\\
		   &= H(U)-\left(H(Z|F)+H(U|Z,F)-H(Z|U,F)\right)\\
		   &= H(U)-\left(H(Z)-H(Z|U,F)\right)\\
		   &\geq H(U)-\left(H(Z|\mathscr{C})+H(U|Z,\mathscr{C})-H(Z|U,\mathscr{C})\right) \label{suff-stat1}\\
		   &= H(U)-H(U|\mathscr{C})\\
		   &= I(U;\mathscr{C}) \label{suff-stat2}
\end{align}
where in (\ref{suff-stat1}) we used the fact that $F$ is a function of $\mathscr{C}$ (not necessarily invertible). In addition, from  $U \rightarrow \mathscr{C} \rightarrow F$, we also have $I(U;F)\leq I(U;\mathscr{C})$, which together with (\ref{suff-stat2}) imply $I(U;F)=I(U;\mathscr{C})$. Thus, $F$ is a sufficient statistics for $\mathscr{C}$ relative to $U$. Therefore, from this point on, we focus on the random cluster formation variable $F$ in our analysis.

The graph model and the resulting cluster formations we described so far are general. For tractability, in this paper, we investigate a specific class of $\mathcal{F}$ which satisfies the following condition: For all $i$, $F_i$ can only be obtained by partitioning some elements of $F_{i-1}$. This assumption results in a tree like structure for cluster formations. Thus, we call $\mathcal{F}$ sets that satisfy this condition \emph{cluster formation trees}. Formally, $\mathcal{F}$ is a cluster formation tree if $F_{i+1}\backslash F_i$ can be obtained by partitioning the elements of $F_{i}\backslash F_{i+1}$ for all $i \in [f-1]$. Note that $\mathcal{F}$ in (\ref{example1-cluster}) is not a cluster formation tree. However, if the probability of the edge between vertices 1 and 3 were 0, then $\mathcal{F}$ would not contain $F_1$ and $F_3$, and $\mathcal{F}$ would be a cluster formation tree in this case. Note that, cluster formation trees may arise in real-life clustering scenarios, for instance, if individuals belong to a hierarchical structure. An example is: an individual may belong to a professor's lab, then to a department, then to a building, then to a campus.  

Next, we define the family of algorithms that we consider, which we coin \emph{two-step sampled group testing algorithms}\footnote{In the two-step sampled group testing algorithms, two steps do not involve consecutive testing phases: the proposed algorithm family in our paper consist of non-adaptive constructions, and should not be confused with semi-adaptive algorithms with two testing phases such as two stage algorithm in \cite{community_aware}.} in this paper. Two-step sampled group testing algorithms consist of two steps in both testing phase and decoding phase. The following definitions are necessary in order to characterize the family of algorithms that we consider in this paper.

In order to design a two-step sampled group testing algorithm, we first pick one of the cluster formations in $\mathcal{F}$ to be the \emph{sampling cluster formation}. The selection of $F_m$ is a design choice, for example, recalling the running example in (\ref{example1})-(\ref{example1-cluster}), one can choose $F_2$ to be the sampling cluster formation.

Next, we define the \emph{sampling function}, $M$, to be a function of $F_m$. The sampling function selects which individuals to be tested by selecting exactly one individual from every subset that forms the partition $F_m$. Let the infected set among the sampled individuals be denoted by $K_M$. The output of the sampling function $M$ is the individuals that are sampled and going to be tested. In the second step, a zero-error non-adaptive group test is performed on the sampled individuals. This results in the identification of the infection status of the selected $\sigma_m = |F_m|$ individuals with zero-error probability. For example, recalling the running example in (\ref{example1})-(\ref{example1-cluster}), when the sampling cluster formation is chosen as $F_2$, we may design $M$ as,
\begin{align} 
	M=\{1,3\}  \label{example1-m2}
\end{align}
Note that, for each selection of $F_m$, $M$ selects exactly 1 individual from each $S_j^m$. As long as it satisfies this property, $M$ can be chosen freely while designing the group testing algorithm.

The \emph{test matrix} $\bm{X}$ is a non-adaptive test matrix of size $T \times \sigma_m$, where $T$ is the required number of tests. Let $U^{(M)}$ denote the infection status vector of the sampled individuals. Then, we have the following test result vector $y$,
\begin{align}
	y_i = \bigvee _ {j \in [\sigma_m]} X_{ij} U_{j}^{(M)}, \quad i \in [T]
\end{align}

In the classical group testing applications, while constructing zero-error non-adaptive test matrices the aim is to obtain unique result vectors, $y$, for every unique possible infected set and for instance, in combinatorial setting, with $d$ infections, \emph{$d$-separable} matrix construction is proposed \cite{hwang_disjunct}. In the classical $d$-separable matrix construction, we have
\begin{align} 
	\bigvee _{i \in S_1} \bm{X}^{(i)} \neq \bigvee _{i \in S_2} \bm{X}^{(i)}\label{f-sep}
\end{align}
for all subsets $S_1$ and $S_2$ of cardinality $d$. As a more general approach, we do not restrict the possible infected sets to the subsets of $[n]$ of the same size, but we consider the problem of designing test matrices that satisfy \eqref{f-sep} for every unique $S_1$ and $S_2$ in a given set of possible infected sets. This approach leads to a more general basis for designing zero-error non-adaptive group testing algorithms for various scenarios, when the set of possible infected sets can be restricted by the available side information.

By using the test result vector $y$, in the first decoding step, the infection status of the sampled individuals are identified with zero-error probability. In the second stage of decoding, depending on $F_m$ and the infection status of the sampled individuals, other non-tested individuals are estimated by assigning the same infection status to all of the individuals that share the same cluster in the cluster formation $F_m$. In the running example, with $M$ given in \eqref{example1-m2}, one must design a zero-error non-adaptive test matrix $\bm{X}$, which identifies the infection status of individuals 1 and 3.

Let $\hat{U} = (\hat{U}_1, \hat{U}_2, \dots , \hat{U}_n)$ be the \emph{estimated infection status vector}. By definition, the infection estimates are the same within each cluster, i.e., for sampling cluster formation $F_m$, $\hat{U}_k = \hat{U}_l$, for all $k,l \in S_j^m$,  for all $j \in [\sigma _m]$. Since $M$ samples exactly one individual from every subset that forms the partition $F_m$, there is exactly one identified individual at the beginning of the second step of the decoding phase and by the aforementioned rule, all $n$ individuals have an estimated infection status at the end of the process. For instance, in the running example, for the sampling cluster formation $F_2$, we have $M=\{1,3\}$ as given in \eqref{example1-m2} and $\bm{X}$ identifies $U_1$ and $U_3$ with zero-error. Then, $\hat{U}_2 = U_1$, since individuals 1 and 2 are in the same cluster in $F_2$. 

Finally, we have two metrics to measure the performance of a group testing algorithm. The first one is the required number of tests $T$, which is the number of rows of $\bm{X}$ in the two step sampled group testing algorithm family that we defined. Having minimum number of required tests is one of the aims of the group testing procedure. The second metric is the expected number of false classifications. Due to the second step of decoding, the overall two step sampled group testing algorithm is not a zero-error algorithm (except for the choice of $m=f$) and the expected number of false classifications is a metric to measure the error performance of the algorithm. We use $E_f = \mathbb{E}[d_H(U \oplus \hat{U})]$ to denote the \emph{expected number of false classifications}, where $d_H(\cdot)$ is the Hamming weight of a binary vector.

Designing a two-step sampled group testing algorithm consists of selecting $F_m$, then designing the function $M$, and then designing the non-adaptive test matrix $\bm{X}$ for the second step of the testing and the first step of the decoding phase for zero-error identification of the infection status of the sampled $\sigma_m$ individuals. We consider cluster formation trees and uniform patient zero assumptions for our infection spread model and we consider two step sampled group testing algorithms for the group test design. 

In the following section, we present a motivating example to demonstrate our key ideas.

\section{Motivating Example}
Consider the following example. There are $n=10$ individuals, and a cluster formation tree with $f=3$ levels. Full characterization of $F$ is as follows,
\begin{align}
	F = 
	\begin{cases}
	F_1=\{\{1,2,3\},\{4,5\},\{6,7,8,9,10\}\}, \enskip &\textit{with probability } 0.4 \\
	F_2=\{\{1,2\},\{3\},\{4,5\},\{6,7,8,9,10\}\}, \enskip &\textit{with probability } 0.2 \\
	F_3=\{\{1,2\},\{3\},\{4,5\},\{6,7\},\{8,9,10\}\}, \enskip &\textit{with probability } 0.4
	\end{cases}
\end{align}
	
First, we find the optimal sampling functions, $M$, for all possible selections of $F_m$. First of all, note that $M$ selects exactly one individual from each subset that forms $F_m$, by definition. Therefore, the number of sampled individuals is constant for a fixed choice of $F_m$. Thus, in the optimal sampling function design, the only parameter that we consider is the minimum number of expected false classifications $E_f$. Note that a false classification occurs only when one of the sampled individuals has a different infection status than one of the individuals in its cluster in $F_m$. For instance, assume that $m=1$ is chosen. Then, assume that the sampling function $M$ selects individual 1 from the set $S_1^1 = \{1,2,3\}$. Recall that after the second step of the two-step group testing algorithm, by using $\bm{X}$, the infection status of individual 1 is identified with zero-error and its status is used to estimate the status of individuals 2 and 3, since they are in the same cluster in $F_m = F_1$. However, with positive probability, individuals 1 and 3 can have distinct infection status, in which case, a false classification occurs. Note that, this scenario occurs only when $F_m$ is at a higher level than the realized $F$ in the cluster formation tree $\mathcal{F}$, where we refer to $F_1$ as the top level of the cluster formation tree and $F_f$ as the bottom level.
	
While finding the optimal sampling function $M$, one must consider the possible false classifications and minimize $E_f$, the expected number of false classifications. As shown in Figure \ref{fig8}, the cluster $\{4,5\}$ does not get partitioned, and for all 3 choices of $F_m$, $M$ can sample either one of the individuals 4 and 5. This selection does not change the expected number of false classifications since $U_4=U_5$ in all possible realizations of $F$. For all sampling cluster formation selections, we have the following analysis:

\begin{figure} [t]
	\centering
	\epsfig{file=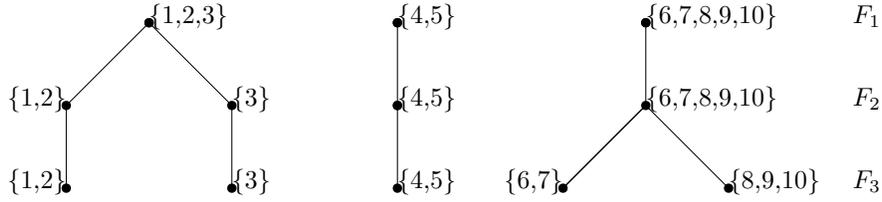,width=.7\textwidth}
	\caption{Cluster formation tree $\mathcal{F}$.}
	\label{fig8}
\end{figure}

\begin{itemize}
    \item \textit{If} $F_m=F_1$:	
	If $M$ samples individual 1 or 2 from the cluster $S_1^1 = \{1,2,3\}$, a false classification occurs if $F=F_2$ and the cluster $\{1,2\}$ is infected, in that case, individual 3 is falsely classified as infected. Similar false classification occurs when $F=F_3$ and the cluster $\{1,2\}$ is infected. Similarly, in these cases, if individual 3 is infected, again, individual 3 is falsely classified as non-infected. Thus, for cluster $\{1,2,3\}$, when either individual 1 or 2 is sampled, the expected number of false classifications is:
	\begin{align} \label{fc1}
		(p_F(F_2)+p_F(F_3))(p_Z(1)+p_Z(2)+p_Z(3)) = 0.6 \times 0.3 = 0.18
	\end{align}
    Similarly, when individual 3 is sampled from the cluster $\{1,2,3\}$, individuals 1 and 2 are falsely classified when $F=F_2$ or $F=F_3$ and either the cluster $\{1,2\}$ or individual 3 is infected. Thus, in that case, the expected number of false classifications is:
    \begin{align} \label{fc2}
		2(p_F(F_2)+p_F(F_3))(p_Z(1)+p_Z(2)+p_Z(3)) = 2 \times 0.6 \times 0.3 = 0.36
	\end{align}
	Thus, (\ref{fc1}) and (\ref{fc2}) imply that, for cluster $S_1^1=\{1,2,3\}$, the optimal $M$ should select either individual 1 or 2 for testing. As discussed above, for cluster $S_2^1=\{4,5\}$, the selection of sampled individual is indifferent and results in 0 expected false classification. Finally, for cluster $S_3^1=\{6,7,8,9,10\}$, a similar analysis implies that, the optimal $M$ should select one of the individuals in $\{8,9,10\}$ for testing.
		
	\item \textit{If} $F_m=F_2$:
	Similar combinatorial arguments follow and we conclude that selection of sampled individuals from the clusters $S_1^2=\{1,2\}$, $S_2^2=\{3\}$ and $S_3^2=\{4,5\}$ are indifferent in terms of the expected number of false classifications. Only possible false classification can happen in cluster $S_4^2=\{6,7,8,9,10\}$ when $F=F_3$ and the infected cluster is either $S_4^3=\{6,7\}$ or $S_5^3=\{8,9,10\}$. Similar to the case $m=1$, if the sampled individual is either 6 or 7, then the expected number of false classifications is 0.6 in contrast to the 0.4 when the sampled individual is one of 8, 9 and 10. Thus, the optimal $M$ should select one of the individuals 8, 9 and 10 as sampled individual to minimize the expected number of false classifications.
	
	\item \textit{If} $F_m=F_3$:
	It is not possible to make a false classification since for all clusters in $F_3$, all individuals that are in the same cluster have the same infection status with probability 1.
\end{itemize}
	
Therefore, for this example, the optimal sampling function selects either individual 1 or 2 from the set $S_1^1$; selects either 4 or 5 from the set $S_2^1$; and selects either 8, 9 or 10 from the set $S_3^1$ if $F_m=F_1$ and the same sampling is optimal with an addition of individual 3, if $F_m=F_2$. Let us assume that $M$ selects the individual with the smallest index when the selection is indifferent among a set of individuals. Thus, the optimal sampling function $M$ for this example is: $\{1,4,8\}$, $\{1,3,4,8\}$ or $\{1,3,4,6,8\}$, depending on the selection of $F_m$ being $F_1$, $F_2$, or $F_3$, respectively. 

Now, for these possible sets of sampled individuals, we need to design zero-error non-adaptive test matrices.

\begin{itemize}
	\item \textit{If} $F_m=F_1$ (i.e., $M = \{1,4,8\}$):
	The set of all possible infected sets is $\mathcal{P}(K_M) = \{\{1\},\{4\},\{8\}\}$. By a counting argument, we need at least 2 tests, since each of three possible infected sets must result in a unique result vector $y$ and each one of these sets has 1 element. We can achieve this lower bound by using the following test matrix:
	\begin{center}
		\begin{tabular}{ c|c|c|c } 
	    	& 1 & 4 & 8 \\ 
			\hline
			Test 1 & 0 & 1 & 1 \\ 
			Test 2 & 1 & 0 & 1 \\ 
		\end{tabular}
	\end{center}
	
	\item \textit{If} $F_m=F_2$ (i.e., $M = \{1,3,4,8\}$): In this case, the set of all possible infected sets is now $\mathcal{P}(K_M) = \{\{1\},\{3\},\{1,3\},\{4\},\{8\}\}$. In the classical zero-error construction for the combinatorial group testing model, one can construct \emph{$d$-separable} matrices, and the rationale behind the construction is to enable the decoding of the infected set, when the infected set can be any $d$-sized subset of $[n]$. However, in our model, the set of all possible infected sets, i.e., $\mathcal{P}(K_M)$, is not a set of all fixed sized subsets of $[n]$, but instead, consists of varying sized subsets of $[n]$ that are structured, depending on the given $\mathcal{F}$. As illustrated in Figure~\ref{fig8}, a given cluster formation tree $\mathcal{F}$ can be represented by a tree structure with nodes\footnote{Throughout the paper, we use the word ``node'' only for the possible clusters in the cluster formation tree representations, not for the vertices in the connection graphs that represent the individuals.} representing possible infected sets, i.e., clusters at each level. Then, the aim of constructing a zero-error test matrix is to have unique test result vectors for each unique possible infected set, i.e., unique nodes in the cluster formation tree. In Figure \ref{fig3}, we present the subtree of $\mathcal{F}$, which ends at the level $F_2$, with assigned result vectors to each node. One must assign unique binary vectors to each node, except for the nodes that do not get partitioned while moving from level to level: those nodes represent the same cluster, and thus, the same vector is assigned, as seen in Figure \ref{fig3}. Moreover, while merging in upper level nodes, binary OR of vectors assigned to the descendant nodes must be assigned to their ancestor node. By combinatorial arguments, one can find the minimum vector length such that such vectors can be assigned to the nodes.
		
	\begin{figure} [t]
		\centering
		\epsfig{file=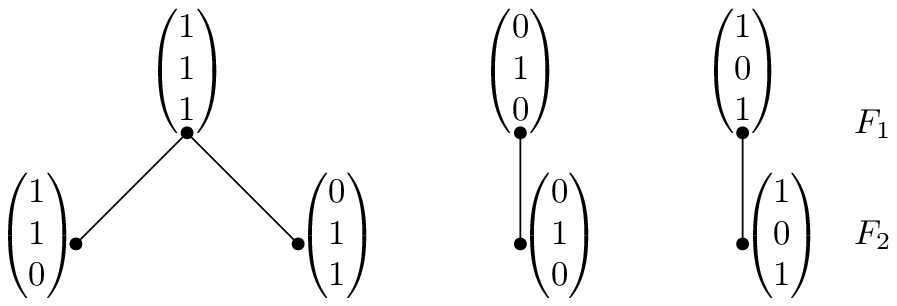,width=0.5\textwidth}
		\caption{Subtree of $\mathcal{F}$ with assigned result vectors for each node.}
		\label{fig3}
	\end{figure}
	
	In this case, the required number of tests must be at least 3 and by assigning result vectors as in Figure \ref{fig3}, we can construct the following test matrix $\bm{X}$:		
	\begin{center}
		\begin{tabular}{ c|c|c|c|c } 
			& 1 & 3 & 4 & 8 \\ 
			\hline
			Test 1 & 1 & 0 & 0 & 1 \\ 
			Test 2 & 1 & 1 & 1 & 0\\ 
			Test 3 & 0 & 1 & 0 & 1
		\end{tabular}
	\end{center}
	Note that for all elements of $\mathcal{P}(K_M)$, corresponding result vector is unique and satisfies the tree structure criteria, as shown in Figure \ref{fig3}.
		
	\item \textit{If} $F_m=F_3$ (i.e., $M = \{1,3,4,6,8\}$):
	In this case, the set of all possible infected sets is $\mathcal{P}(K_M)=\{\{1\},\{3\},\{1,3\},\{4\},\{6\},\{8\},\{6,8\}\}$. We give a tree structure representation with assigned result vectors of length 3 that achieves the tree structure criteria discussed above, which is shown in Figure~\ref{fig4} where each unique node is assigned a unique vector except for the nodes that do not get partitioned while moving from level to level. Note that every unique node in the tree representation corresponds to a unique element of $\mathcal{P}(K_M)$. The corresponding test matrix $\bm{X}$ is the following $3 \times 5$ matrix:
	\begin{center}
		\begin{tabular}{ c|c|c|c|c|c } 
			& 1 & 3 & 4 & 6 & 8 \\ 
			\hline
			Test 1 & 1 & 0 & 0 & 1 & 0 \\ 
			Test 2 & 1 & 1 & 1 & 0 & 0\\ 
			Test 3 & 0 & 1 & 0 & 0 & 1
		\end{tabular}
	\end{center}
		
	\begin{figure} [t]
		\centering
		\epsfig{file=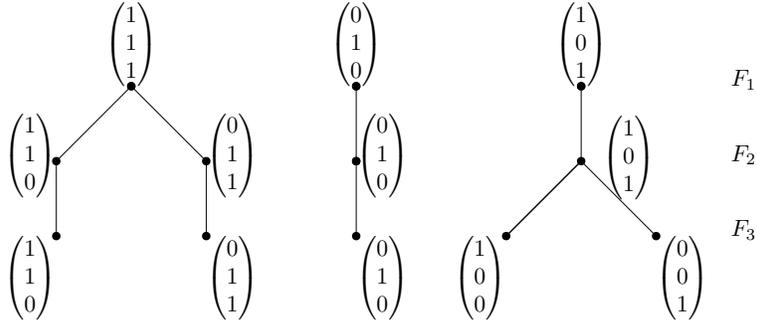,width=.6\textwidth}
		\caption{$\mathcal{F}$ with assigned result vectors for each node.}
		\label{fig4}
	\end{figure}	
\end{itemize}
	
A more structured and detailed analysis of the selection of the optimal sampling function and the minimum number of required tests is given in the next section.
	
We finalize our analysis of this example by calculating the expected number of false classifications where $E_{f,\alpha}$ denotes the conditional expected false classifications, given $F=F_\alpha$:

\begin{itemize}
	\item \textit{If} $F_m=F_1$:
	\begin{align}
	   	E_f &= \sum_{\alpha}p_F(F_\alpha)E_{f,\alpha} \nonumber \\ 
				&= p_F(F_2)E_{f,2} + p_F(F_3)E_{f,3} \nonumber \\
				&= 0.2(0.3 \times 1) + 0.4(0.3 \times 1 + 0.5 \times 2) \nonumber \\
				&= 0.58
	   \end{align}
	\item \textit{If} $F_m=F_2$:
	\begin{align}
	   	E_f 	&= p_F(F_3)E_{f,3} \nonumber \\
				&= 0.4(0.5 \times 2) \nonumber \\
				&= 0.4
	\end{align}
	\item \textit{If} $F_m=F_3$, we have $E_f = 0$.
\end{itemize}
	
Note that the choice of $F_m$ is a design choice and one can use time sharing\footnote{Time sharing can be implemented by assigning a probability distribution to $F_m$ over $\mathcal{F}$, instead of picking one cluster formation from $\mathcal{F}$ to be $F_m$ deterministically.} between different choices of $m$, depending on the specifications of the desired group testing algorithm. For instance, if a minimum number of tests is desired, then one can pick $m=1$, which results in 2 tests, which is the minimum possible, but with expected 0.58 false classifications, which is the maximum possible in this example. On the other hand, if a minimum expected false classifications is desired, then one can pick $m=3$, results in 0 expected false classifications, which is the minimum possible, but with 3 tests, which is the maximum possible in this example. Generally, there is a trade-off between the number of tests and the number of false classifications, and we can formulate optimization problems for specific system requirements, such as finding a time sharing distribution for $F_m$ that minimizes the number of tests for a desired level of false classifications, or vice versa. 

In the following section, we describe the details of our proposed group testing algorithm.

\section{Proposed Algorithm and Analysis} \label{sec4}
In our $\mathcal{F}$-\textit{separable matrix} construction, we aim to construct binary matrices that have $n$ columns, and for each possible infected subset of the selected individuals, there must be a corresponding distinct result vector. A binary matrix $\bm{X}$ is $\mathcal{F}$-separable if
\begin{align} 
	\bigvee _{i \in S_1} \bm{X}^{(i)} \neq \bigvee _{i \in S_2} \bm{X}^{(i)}\label{d-sep}
\end{align}
is satisfied for all distinct subsets $S_1$ and $S_2$ in the set of all possible infected subsets, where $\bm{X}^{(i)}$ denotes the $i${th} column of $\bm{X}$. In $d$-separable matrix construction \cite{hwang_disjunct}, this condition must hold for all subsets $S_1$ and $S_2$ of cardinality $d$; here, it must hold for all possible feasible infected subsets as defined by $\mathcal{F}$. From this point of view, our $\mathcal{F}$-separable test matrix construction exploits the known structure of $\mathcal{F}$ and thus, it results in an efficient zero-error non-adaptive test design for the second step of our proposed algorithm.

We adopt a combinatorial approach to the design of the non-adaptive test matrix $\bm{X}$. Note that, for a given $M$, we have $\sigma_m$ individuals to be identified with zero-error probability. The key point of our algorithm is the fact that the infected set of individuals among those selected individuals can only be some specific subsets of those $\sigma_m$ individuals. Without any information about the cluster formation, any one of the $2^{\sigma_m}$ subsets of the selected individuals can be the infected set. However, since we are given $\mathcal{F}$, we know that, the infected set among the selected individuals, $K_M$, can be one of the $2^{\sigma_m}$ subsets only if there exists at least one set $S_i^j$ that  contains $K_M$ and there is no element in the difference set $M\backslash K_M$ such that it is an element of all sets $S_i^j$ containing $K_M$. This fact, especially in a cluster formation tree structure, significantly reduces the total number of possible infected subsets that need to be considered. Therefore, we can focus on such subsets and design the test matrix $\bm{X}$ by requiring that the logical OR operation of the columns that correspond to the possible $K_M$ sets to be distinct, in order to decode the test results with zero-error. Let $\mathcal{P}(K_M)$ denote the set of possible infected subsets of the selected individuals, i.e., the set of possible sets that $K_M$ can be. Then, matrix $\bm{X}$ must satisfy \eqref{d-sep} for all distinct $S_1$ and $S_2$ that are elements of $\mathcal{P}(K_M)$. Note that, the decoding process is a mapping from the result vectors to the infected sets and thus, we require the distinct result vector property to guarantee zero-error decoding.

Designing the $\bm{X}$ matrix that satisfies the aforementioned property is the key idea of our algorithm. Before going into the design of $\bm{X}$, we first derive the expected number of false classifications in a given two step sampled group testing algorithm. Recall that false classifications occur during the second step of the decoding phase. In particular, in the second step of the decoding phase, depending on the selection of the sampling cluster formation $F_m$, the infection status of selected individuals $M$ are assigned to the other individuals such that the infection status estimate is the same within each cluster. For fixed sampling cluster formation $F_m$ and the sampling function $M$, the number of expected false classifications can be calculated as in the following theorem.

\begin{theorem}\thlabel{theorem1}
	In a two step sampled group testing algorithm with the given sampling cluster formation $F_m$ and the sampling function $M$ over a cluster formation tree structure defined by $\mathcal{F}$ and $p_F$, with uniform patient zero distribution $p_Z$ over $[n]$, the expected number of false classifications given $F = F_\alpha$ is
	\begin{align}
	E_{f,\alpha} = \sum_{i \in [\sigma_m]} \left(\frac{|S^\alpha(M_i)|}{n} \cdot |S_i^m \backslash S^\alpha(M_i)|+\sum_{S_j^\alpha \subseteq S_i^m \backslash S^\alpha(M_i)}\frac{|S_j^\alpha|^2}{n}\right) \label{condfalse}
	\end{align}
	and the expected number of false classifications is
	\begin{align}
	E_f = \sum_{\alpha >m}p_F(F_\alpha)E_{f,\alpha} \label{falseclass}
	\end{align}
	where $S^\alpha(M_i)$ is the subset in the partition $F_\alpha$ which contains the $i$th selected individual.
\end{theorem}

\begin{Proof}
	For the sake of simplicity, we denote the subset in partition $F_\alpha$ that contains the $i$th selected individual by $S^\alpha(M_i)$. We start our calculation with the conditional expectation where $F = F_\alpha$ is given. Observe that an error occurs, in the second step of the decoding process, only if $F_m$ is at a higher level of the cluster formation tree than the realization of $F = F_\alpha$ and the true infected cluster $K = S_\gamma^\alpha$ is merged at the level $F_m$, i.e., $\alpha>m$ and $S_\gamma^\alpha \notin F_m$. Since there is exactly one true infected cluster, which is at level $F_\alpha$, false classifications only happen in the set $S_\theta^m$ that contains $S_\gamma^\alpha$. Now, we know that for the given sampling function $M$, the $\theta$th selected individual is selected from the set $S_\theta^m$ and in the second step of the decoding phase, its infection status is assigned to all of the members of the set $S_\theta^m$. Therefore, the members of the difference set $S_\theta^m \backslash S^\alpha(M_\theta)$ are falsely classified if the set $S^\alpha(M_\theta)$ is the true infected set. In that case, all members of $S_\theta^m$ would be classified as infected while only the subset of them, which is $S^\alpha(M_\theta)$ were infected. On the other hand, when the cluster of the selected individual at level $F_\alpha$ is not infected, i.e., the infected cluster is a subset of $S_\theta^m \backslash S^\alpha(M_\theta)$, then only the infected cluster is falsely identified, since all of the members of $S_\theta^m$ are classified as non-infected. Thus, we have the following conditional expected number of false classifications when $F = F_\alpha$ is given, where $p_{S_i^j}$ denotes the probability of the set $S_i^j$ being infected	
	\begin{align}
		E_{f,\alpha} &= \sum_{i \in [\sigma_m]} \left(p_{S_{M_i}^\alpha}|S_i^m \backslash S^\alpha(M_i))|+\sum_{S_j^\alpha \subseteq S_i^m \backslash S^\alpha(M_i)}p_{S_j^\alpha}|S_j^\alpha|\right) \\		
		&= \sum_{i \in [\sigma_m]} \left(\frac{|S^\alpha(M_i)|}{n} \cdot |S_i^m \backslash S^\alpha(M_i)|+\sum_{S_j^\alpha \subseteq S_i^m \backslash S^\alpha(M_i)}\frac{|S_j^\alpha|^2}{n}\right) \label{conderror}
	\end{align} 
	where \eqref{conderror} follows from the uniform patient zero assumption. Finally, since false classifications occur only when $\alpha > m$, we have the following expression for the expected number of false classifications
	\begin{align}
		E_f = \sum_{\alpha > m}p_F(F_\alpha)E_{f,\alpha}
	\end{align} 
	concluding the proof.
\end{Proof}

We now proceed to characterize the optimal choice of the sampling function $M$. First, we define $\beta_i(k)$ functions as follows. For $i \in [f]$ and $k \in [n]$,
\begin{align}
	\beta_i(k) \triangleq \sum_{j>i} p_F(F_j) \left(|S^j(k)| \cdot |S^i(k) \backslash S^j(k)|+\sum_{S_l^j \subseteq S^i(k) \backslash S^j(k)}|S_l^j|^2\right) \label{beta}
\end{align}
where $S^i(k)$ is the subset in partition $F_i$ that contains $k$.

\begin{theorem}\thlabel{lemma1}
	For sampling cluster formation $F_m$, the optimal choice of $M$ that minimizes the expected number of false classifications is
	\begin{align}
		M_i = \underset{k \in S_i^m}{\arg\min} ~ \beta_m(k) \label{optsample}
	\end{align}
	where $M_i$ is the $i$th selected individual. Moreover, the number of required tests is constant and is independent of the choice of $M$.
\end{theorem}

\begin{Proof}
	We first prove the second part of the theorem, i.e., that the choice of $M$ does not change the required number of tests. In a cluster formation tree structure, when we sample exactly one individual from each subset $S_i^m$, $\mathcal{P}(K_M)$ contains single element subsets of selected individuals, since when $F = F_m$ we have exactly one infected individual that can be any one of these individuals with positive probability. Now consider the cluster formation $F_{m-1}$. Since it is a cluster formation tree structure, there must be at least one $S_i^{m-1}$ such that, $S_i^{m-1} = S_j^m \cup S_k^m, ~ S_j^m \neq S_k^m$, which means that, $\mathcal{P}(K_M)$ must contain the set of selected individuals from $S_k^m$ and $S_j^m$ as well, because of the fact that in the case of $F=F_{m-1}$, these individuals can be infected simultaneously. Similarly, when moving towards the top node of the cluster formation tree (i.e., $F_1$), whenever we observe a merging, we must add corresponding union of the subsets of individuals to $\mathcal{P}(K_M)$, which is the set of all possible infected sets for the selected individuals $M$. Thus, the structure of distinct sets of possible infected individuals do not depend on the indices of the sampled individuals within each $S_i^m$, but depends on the given $\mathcal{F}$ and $F_m$, completing the proof of the second part of the theorem.
	
	We next prove the first part of the theorem, i.e., we prove that selecting the individual that has the minimum $\beta_m(k)$ value for each $S_i^m$ results in the minimum expected number of false classifications and thus, it is the optimal choice. First, recall that, by definition, $M$ depends on $F_m$ and thus, we design sampling function $M$ for a given $F_m$. Now, recall the expected number of false classifications stated in \eqref{condfalse}-\eqref{falseclass}. Designing a sampling function that minimizes $E_f$ for a given $F_m$ can be done as follows. From \eqref{condfalse}-\eqref{falseclass},
	\begin{align}
		\underset{M}{\min} ~ E_f &= \underset{M}{\min}  \left\{\sum_{\alpha: m<\alpha}p_F(F_\alpha)\sum_{i \in [\sigma_m]} \left(\frac{|S^\alpha(M_i)|}{n} \cdot |S_i^m \backslash S^\alpha(M_i)|+\sum_{S_j^\alpha \subseteq S_i^m \backslash S^\alpha(M_i)}\frac{|S_j^\alpha|^2}{n}\right)\right\} \\
		&=\frac{1}{n}\sum_{i \in [\sigma_m]}\underset{M}{\min}  \left\{\sum_{\alpha: m<\alpha}p_F(F_\alpha) \left(|S^\alpha(M_i)| \cdot |S_i^m \backslash S^\alpha(M_i)|+\sum_{S_j^\alpha \subseteq S_i^m \backslash S^\alpha(M_i)}|S_j^\alpha|^2\right)\right\}\\
		&=\frac{1}{n}\sum_{i \in [\sigma_m]}\left(\sum_{\alpha: m<\alpha}p_F(F_\alpha)\left(|S^\alpha(k_i^*)| \cdot |S_i^m \backslash S^\alpha(k_i^*)|+ \sum_{S_j^\alpha \subseteq S_i^m \backslash S^\alpha(k_i^*)}|S_j^\alpha|^2\right)\right) \label{minexpfalse}
	\end{align}
	where $k_i^* = \underset{k \in S_i^m}{\arg\min} ~ \beta_m(k)$, and \eqref{minexpfalse} is the minimum value of the expected number of false classifications for given $F_m$. The sampling function $M$ defined in \eqref{optsample} achieves the minimum and thus, it is optimal, completing the proof of the first part of the theorem.
\end{Proof}

The optimal $M$ analysis focuses on choosing the sampling function that results in the minimum expected number of false classifications, among the set of functions that select exactly one individual from each cluster of a given $F_m$. For some scenarios, it is possible to choose a sampling function that selects multiple individuals from some clusters of a given $F_m$ that achieves expected false classifications-required number of tests points that cannot be achieved by the optimal $M$ in \eqref{optsample}. However, majority of the cases, the sampling functions of interest, i.e., the sampling functions that choose exactly one individual from each $F_m$, are globally optimal. First, the sampling functions that select multiple individuals from a cluster that never gets partitioned further in the levels below $F_m$ is sub-optimal: these sampling functions select multiple individuals to identify who are guaranteed to have the same infection status. For instance, in zero expected false classifications case, i.e., the bottom level $F_f$ is chosen as the sampling cluster formation, sampling more than one individual from each cluster is sub-optimal. Second, picking the sampling cluster formation $F_m$ and choosing an $M$ such that multiple individuals are chosen from some clusters that further get partitioned in the levels below $F_m$, is equivalent to choosing a sampling cluster formation below $F_m$ and using an $M$ that selects exactly one individual from each cluster of the new sampling cluster formation, except for the scenarios where there exist partitioning of multiple clusters in two consecutive cluster formations in a given $\mathcal{F}$, and one can consider a sampling function that selects multiple individuals from some clusters of a given $F_m$ that cannot be represented as a sampling function that selects exactly one individual from each cluster of another cluster formation $F_{m^\prime}$. For the sake of compactness, we focus on the family of sampling functions $M$ that selects exactly one individual from each cluster of the chosen $F_m$.

So far we have presented a method to select individuals to be tested in a way to minimize the expected number of false classifications. Now, we move on to the design of $\bm{X}$, the zero-error non-adaptive test matrix which identifies the infection status of the selected individuals $M$ with a minimum number of tests. Recall that since $|\mathcal{F}|=f$, there are $f$ possible choices of $F_m$ and each choice results in a different test matrix $\bm{X}$.

Based on the combinatorial viewpoint stated in \eqref{d-sep}, we propose a family of non-adaptive group testing algorithms which satisfy the separability condition for all of the subsets in $\mathcal{P}(K_M)$, which is determined by $\mathcal{F}$. We call such matrices \textit{$\mathcal{F}$-separable matrices} and non-adaptive group tests that use \textit{$\mathcal{F}$-separable matrices} as their test matrix as \textit{$\mathcal{F}$-separable non-adaptive group tests}. In the rest of the section, we present our results on the required number of tests for \textit{$\mathcal{F}$-separable non-adaptive group tests}.

The key idea of designing an $\mathcal{F}$-separable matrix is determining the set $\mathcal{P}(K_M)$ for a given set of selected individuals $M$ and the tree structure of $\mathcal{F}$ so that we can find binary column vectors for each selected individual where all of the corresponding possible result vectors are distinct. Note that, for a given choice of $F_m$, if we consider the corresponding subtree of $\mathcal{F}$ which starts from the first level $F_1$ and ends at the level $F_m$, the problem of finding an $\mathcal{F}$-separable non-adaptive test matrix is equivalent to finding a set of length $T$ binary column vectors for each node at level $F_m$ that satisfy the following criteria:
\begin{enumerate}
	\item For every node at the levels that are above the level $F_m$, each node must be assigned a binary column vector that is equal to the OR of all vectors that are assigned to its descendant nodes. This is because each node in the tree corresponds to a possible set of infected individuals among the selected individuals where each merging of the nodes corresponds to the union of the possible infected sets which results in taking the OR of the assigned vectors of the merged nodes. 
	\item Each assigned binary vector must be unique for each unique node, i.e., for every node that represents a unique set $S_i^j$. For the nodes that do not split between two levels, assigned vector remains the same. This is because each unique node (note that when a node does not split between levels, it still represents the same set of individuals) corresponds to a unique possible infected subset of the selected individuals and they must satisfy \eqref{d-sep}.
\end{enumerate}
In other words, for a cluster formation tree with assigned result vectors to each node, a sufficient condition for achievability of $\mathcal{F}$-separable matrices as follows:
\begin{itemize}
\item [] Let $u$ be a node with Hamming weight $d_H(u)$. Then, the number of all descendant nodes of $u$ with constant Hamming weights $i$ must be less than $\binom{d_H(u)}{i}$ for all $i$. This must hold for all nodes $u$. Furthermore, number of nodes with constant Hamming weight $i$ must be less than $\binom{T}{i}$ for all $i$. In addition, Hamming weights of the nodes must strictly decrease while moving from ancestor nodes to descendant nodes.
\end{itemize}
This condition is indeed sufficient because it guarantees the existence of unique set of vectors that can be assigned to each node of the subtree of $\mathcal{F}$ that satisfies the merging/OR structure determined by the subtree. 

The problem of designing an $\mathcal{F}$-separable non-adaptive group test can be reduced to finding the minimum number $T$, for which we can find $\sigma_m$ binary vectors with length $T$, such that all vectors that are assigned to the nodes satisfy the above condition. Here the assigned vectors are the result vectors $y$ when the corresponding node is the infected node.

We have the following definitions that we need in \thref{lemma2}. For a given $\mathcal{F}$, we define $\lambda_{S_i^j}$ as the number of unique ancestor nodes of the set $S_i^j$. We also define $\lambda_{j}$ as the number of unique sets $S_a^b$ in $\mathcal{F}$ at and above the level $F_j$. Note that $\sum _{a\leq j} \sigma_a$ is the total number of sets $S_a^b$ in $\mathcal{F}$ at and above the level $F_j$, and thus we have,
\begin{align}
	\sum_{a\leq j} \sigma_a \geq \lambda_{j}
\end{align}

\begin{theorem}\thlabel{lemma2}
	For given $\mathcal{F}$ and $F_m$ for $m<f$, the number of required tests for an \textit{$\mathcal{F}$-separable non-adaptive group test}, i.e., the number of rows of the test matrix $\bm{X}$, must satisfy
	\begin{align} \label{lemma2-eqn}
		T \geq \max\left\{\underset{j \in [\sigma_m]}{\max}~(\lambda_{S_j^m}+1), \enskip \ceil*{\log_2(\lambda_{m}+1)}\right\}
	\end{align}
	with addition of 1's removed in (\ref{lemma2-eqn}) for the special case of $m=f$. 
\end{theorem}

\begin{Proof}
	First, we have that each unique node (nodes that represent a unique subset $S_i^j$) represents a unique possibly infected set $K_M$ where each result vector must be unique as well. Therefore, in total, we must have at least $\lambda_m$ unique vectors. Furthermore, when $m<f$, it is possible that the infected set among the sampled individuals is the empty set. Thus, we have to reserve the zero vector for this case as well. Therefore, the total number of tests must be at least $\ceil*{\log_2(\lambda_{m}+1)}$ in general, with an exception of $m=f$ case, where we can assign the zero vector to one of the nodes and may achieve $\ceil*{\log_2(\lambda_{m})}$.
	
	Second, assume that for any node $j$ at an arbitrary level $F_i$, $i<m$, the set of indices of the positions of 1's must contain the set of indices of the positions of 1's of the descendants of node $j$. Moreover, since all nodes that split must be assigned a unique vector, Hamming weights of the vectors must strictly decrease as we move from an ancestor node to a descendant at each level. Considering the fact that the ancestor node at the top level can have Hamming weight at most $T$ and the nodes at the level $F_m$ must be assigned a vector which has Hamming weight at least 1, including the node that has the most unique ancestor nodes, $T$ must be at least $\underset{j \in [\sigma_m]}{\max}(\lambda_{S_j^m}+1)$. Similar to the first case, when $m=f$, we can have zero vector assigned to one of the bottom level nodes and thus, we can have $T$ at least $\underset{j \in [\sigma_m]}{\max}\lambda_{S_j^m}$.
\end{Proof}

Note that \thref{lemma2} is a converse argument, without a statement about the achievability of the given lower bound. In fact, the given lower bound is not always achievable. 

\paragraph{Complexity:}The time complexity of the two-step sampled group testing algorithms consists of the complexity of finding the optimal $M$ given $F_m$ and $\mathcal{F}$, the complexity of the construction of the $\mathcal{F}$-separable test matrix given $M$ and $\mathcal{F}$, and the complexity of the decoding of the test results given the test matrix $\bm{X}$ and the result vector $y$. In the following lemmas, we analyze the complexity of these processes.

\begin{lemma} \label{complexity1}
    For a given cluster formation tree $\mathcal{F}$ and a sampling cluster formation $F_m$, the complexity of finding the optimal $M$ as in \thref{lemma1} is
    \begin{align}
        O(n(f-m)\zeta_m)
    \end{align}
    where $\zeta_m = \max \limits_{k \in [n]}|\{S_l^f: S_l^f \subseteq S^m(k) \backslash S^f(k)\}|$. 
\end{lemma}

\begin{Proof}
    In order to find the optimal $M$, $\beta_m(k)$ needs to be calculated as in \eqref{beta} for each $k \in [n]$. The complexity of each of these calculations is bounded above by the number of cluster formations below $F_m$ multiplied by the number of clusters at level $f$, that do not include the individual $k$ and form the cluster $S^m(k)$, i.e., the clusters $S_l^f$ that satisfy $S_l^f \subseteq S^m(k) \backslash S^f(k)$. Note that this upper bound varies for each $k \in [n]$ and the total complexity is the summation of these sizes multiplied by $f-m$, i.e., the number of cluster formations below $F_m$, for each $k \in [n]$. As an upper bound, we consider the maximum of these sizes, i.e., $\zeta_m$, concluding the proof.
\end{Proof}

In the next lemma, we analyze the complexity of the construction of the $\mathcal{F}$-separable test matrix given $M$ and $\mathcal{F}$.

\begin{lemma} \label{complexity2}
    For a given cluster formation tree $\mathcal{F}$ and a sampling function $M$, the complexity of assigning the binary result vectors to the nodes in $\mathcal{F}$, and thus, the construction of the $\mathcal{F}$-separable test matrix is $\Omega(m\sigma_m)$.
\end{lemma}

\begin{Proof}
    When the cluster formation tree $\mathcal{F}$ and the sampling function $M$ are given, in order to assign unique binary result vectors to each node in $\mathcal{F}$ that represents a unique possible infected cluster, we need to consider the subtree of $\mathcal{F}$ that starts with the level $F_1$ and ends at the level $F_m$, as in the example in Figure~\ref{fig3}. Then, we need to traverse from each bottom node in the subtree, to the top node, to detect every merging of each cluster. This results in finding the numbers $\lambda _{S_j^m}$ for $j \in [\sigma_m]$ and $\lambda _m$ and unique binary test result vectors can be assigned to each unique node in $\mathcal{F}$. The traversing on the subtree of $\mathcal{F}$ starting from the bottom level $F_m$ to the top level for each bottom level node has the complexity $\Theta(m\sigma_m)$. This traversing does not immediately result in the explicit construction of unique binary result vectors to be assigned, but it gives an asymptotic lower bound for the complexity of the construction of the $\mathcal{F}$-separable test matrices.
\end{Proof}

Note that the Lemma~\ref{complexity2} is an asymptotic lower bound for the complexity of the binary result vector assignment to the unique nodes in $\mathcal{F}$, and thus, for the construction of the $\mathcal{F}$-separable test result matrix $\bm{X}$. This analysis is a baseline for the proposed model and proposing explicit $\mathcal{F}$-separable test matrix constructions with exact number of required tests and complexity is an open problem.

\begin{lemma}\label{complexity3}
    For a given $\mathcal{F}$-separable test matrix $\bm{X}$, with corresponding cluster formation tree $\mathcal{F}$ with assigned binary result vectors to each node and the result vector $y$, the decoding complexity is $O(1)$.
\end{lemma}

\begin{Proof}
    While constructing the $\mathcal{F}$-separable test matrix, we consider the assignment of the unique binary result vectors to the nodes in the given cluster formation tree $\mathcal{F}$. For a given test matrix $\bm{X}$ and the result vector $y$, the decoding problem is a hash table lookup, with the complexity $O(1)$.
\end{Proof}

Since during the proposed process of assignment of unique binary result vectors to each unique node in $\mathcal{F}$, we specifically assign the test result vectors to every unique possible infected set, the decoding process is basically a hash table lookup, resulting in fast decoding with low complexity.

In the next section, we introduce and focus on a family of cluster formation trees which we call \textit{exponentially split cluster formation trees}. For this analytically tractable family of cluster formation trees, we achieve the lower bound in \thref{lemma2}  \emph{order-wise}, and we compare our result with the results in the literature. 

\section{Exponentially Split Cluster Formation Trees} \label{sec5}
In this section, we consider a family of cluster formation trees, explicitly characterize the selection of optimal sampling function, and the resulting expected number of false classifications and the number of required tests. We also compare our results with Hwang's generalized binary splitting algorithm \cite{hwang_binary} and zero-error non-adaptive group testing algorithms in order to show the gain of utilizing the cluster formation structure as done in this paper.

A cluster formation tree $\mathcal{F}$ is an \emph{exponentially split cluster formation tree} if it satisfies the following criteria:
\begin{itemize}
	\item An exponentially split cluster formation tree that consists of $f$ levels has $2^{i-1}$ nodes at level $F_i$, for each $i \in [f]$, i.e., $\sigma_i = 2^{i-1},  i \in [f]$.
	\item At level $F_i$, every node has $2^{f-i}\delta$ individuals where $\delta$ is a constant positive integer, i.e., $|S_j^i| = 2^{f-i}\delta,  i \in [f],  j \in [\sigma_i]$.
	\item Every node has exactly 2 descendant nodes in one level below in the cluster formation tree, i.e., every node is partitioned into equal sized 2 nodes when moving one level down in the cluster formation tree.
	\item Random cluster formation variable $F$ is uniformly distributed over $\mathcal{F}$, i.e., $p_F(F_i) = 1/f, i \in [f]$.
\end{itemize}

\begin{figure} [t]
	\centering
	\epsfig{file=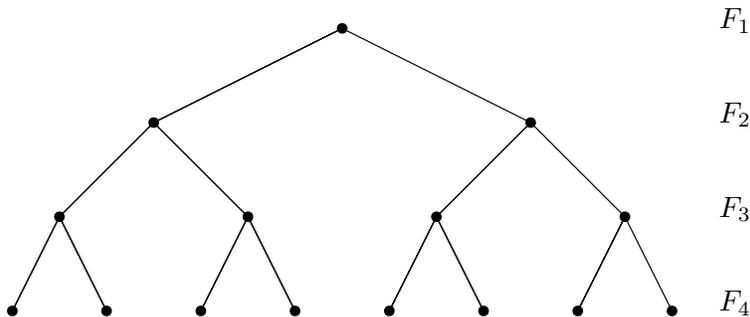,width=.6\textwidth}
	\caption{A 4 level exponentially split cluster formation tree.}
	\label{fig6}
\end{figure}

We analyze the expected number of false classifications and the required number of tests for exponentially split cluster formation trees, by using the general results derived in Section~\ref{sec4}. In Figure~\ref{fig6}, we give a 4 level exponentially split cluster formation tree example. In that example, there is $2^0 = 1$ node at level $F_1$ and the number of nodes gets doubled at each level, since each node is split into two nodes when moving one level down in the tree. Also, the sizes of the nodes that are at the same level are the same, with the bottom level nodes having the size $\delta$.

\begin{figure} [t] 
	\centering
	\begin{subfigure}[b]{0.475\textwidth}
		\centering
		\epsfig{file=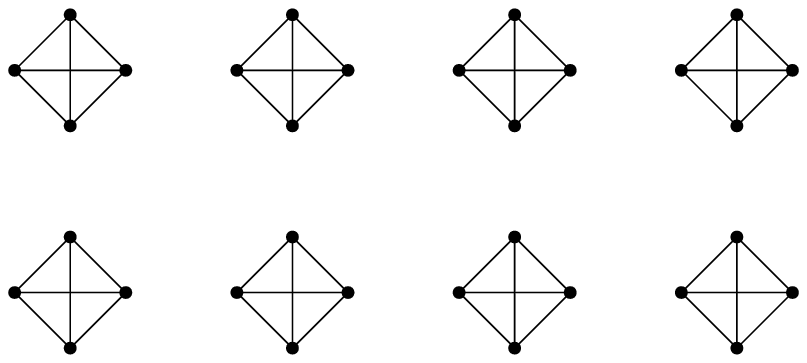,width=.75\textwidth}
		\caption[]%
		{{\small $F_4$}}    
	\end{subfigure}
	\hfill
	\begin{subfigure}[b]{0.475\textwidth}  
		\centering 
		\epsfig{file=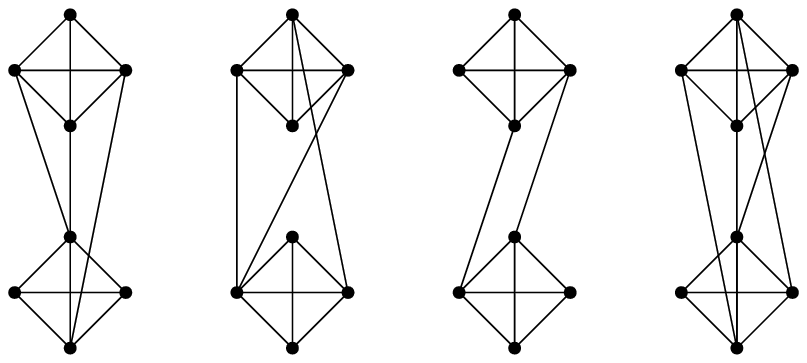,width=.75\textwidth}
		\caption[]%
		{{\small $F_3$}}
	\end{subfigure}
	\vskip\baselineskip
	\begin{subfigure}[b]{0.475\textwidth}   
		\centering 
		\epsfig{file=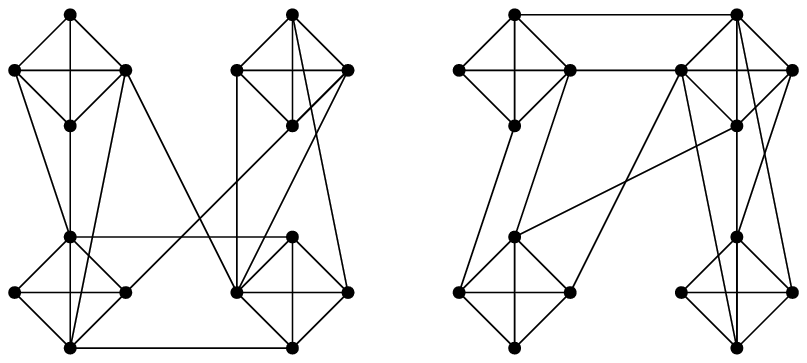,width=.75\textwidth}
		\caption[]%
		{{\small $F_2$}}
	\end{subfigure}
	\hfill
	\begin{subfigure}[b]{0.475\textwidth}   
		\centering 
		\epsfig{file=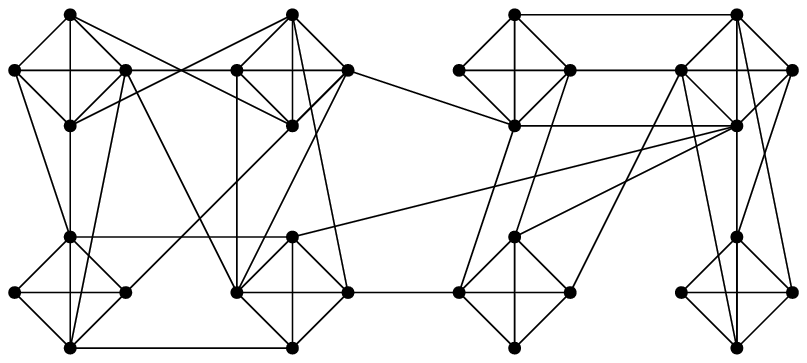,width=.75\textwidth}
		\caption[]%
		{{\small $F_1$}}
	\end{subfigure}
	\caption{4 realizations of a random connection graph $\mathcal{C}$ that falls under 4 different cluster formations in a 4-level exponentially split cluster formation tree with $\delta = 4$.}
	\label{fig6extra}
\end{figure}

Being a subset of cluster formation trees, exponentially split cluster formation trees correspond to random connection graphs where edges between individuals are not independently realized in non-trivial cases. For instance, in Figure~\ref{fig6extra}, we present 4 different possible realizations of edges of a 4-level exponentially split cluster formation tree system, given in Figure~\ref{fig6}, where there are $\delta = 4$ individuals in the bottom level clusters. Here, if the edges between individuals are realized independently, then there would be possible cluster formations that do not result in exponentially split cluster formation tree structure. The edge realizations are correlated in the sense that, if there is at least one edge realized between two bottom level neighbor clusters, then there must be at least one edge realized between other bottom level neighbor cluster pairs as well. Similarly, if there is at least one bottom level cluster pair that are not immediate neighbors but get merged in some upper level $F_k$ in $\mathcal{F}$, then other bottom level cluster pairs that get merged in $F_k$ must be connected as well. In Figure~\ref{fig6extra}, in $F_4$ realization, the only edges that are present are the edges that form bottom level clusters. In $F_3$ realization, there are at least one edge realized between each bottom level neighbor cluster pair, resulting in clusters of 8 individuals. Similarly, there are more distant connections that are realized in $F_2$ and $F_1$. From a practical point of view, the 4-level exponential split cluster formation tree example in Figure~\ref{fig6} and Figure~\ref{fig6extra}, can be used to model real-life scenarios, such as the infection spread in an apartment complex with multiple buildings. In the bottom level, there are households that are guaranteed to be connected, and in $F_3$ level the households that are in close contact are connected, in $F_2$ level there is a connection building-wise and in $F_1$ the whole community is connected. Note that, the connections given in Figure~\ref{fig6extra} are realization examples that fall under 4 possible cluster formations and all edge realization scenarios are possible as long as the resulting cluster formation is one of the four given cluster formations. While designing the group testing algorithm, the given information is the probability distribution over the cluster formations, and in practice, one can expect a probability distribution where bottom level cluster formations, i.e., cluster formations towards $F_4$, have higher probabilities in a community where there are strict social isolation measures, and high immunity rates for a contagious infection whereas higher probabilities of upper level cluster formations, i.e., cluster formations toward $F_1$, can be expected for communities with high contact rate and lower immunity.

\paragraph{Optimal sampling function and expected number of false classifications:} Due to the symmetry of the system, for any choice $F_m$, each element of $S_i^m$ has the same $\beta_m(i)$ value for all $i \in \sigma_m$. Therefore, the sampling function selects individuals from each set arbitrarily, i.e., the selection of a particular individual does not change the expected number of false classifications. Thus, we can pick any sampling function that selects one element from each $S_i^m$. By \thref{theorem1}, the expected number of false classifications, for given $F_m$, is 
\begin{align}
	E_f &= \sum_{\alpha > m}\frac{1}{f}\sum_{i \in [\sigma_m]} \left(\frac{|S^\alpha(M_i)|}{n} \cdot |S_i^m \backslash S^\alpha(M_i)|+\sum_{S_j^\alpha \subseteq S_i^m \backslash S^\alpha(M_i)}\frac{|S_j^\alpha|^2}{n}\right) \\
	&= \sum_{\alpha > m}\frac{1}{f}\frac{\sigma_m}{\sigma_\alpha}\left(\delta(2^{f-m}-2^{f-\alpha})+(2^{\alpha-m}-1)\delta 2^{f-\alpha} \right)\\
	&=\sum_{\alpha > m}\frac{2^{f+1}\delta}{f}\left(2^{-\alpha}-2^{m-2\alpha} \right)\\
	&=\frac{2^{f+1}\delta}{f}\left(\sum_{\alpha > m}2^{-\alpha}-2^{m}\sum_{\alpha > m}2^{-2\alpha}\right) \\
	&=\frac{2^{f+1}\delta}{f}\left((2^{-m}-2^{-f})-\frac{2^{m}}{3}(2^{-2m}-2^{-2f})\right)\\
	&=\frac{\delta}{3f}\left(2^{f-m+2}+2^{m-f+1}-6\right)
\end{align}
This expected number of false classifications takes its maximum value when $F_m = F_1$,
\begin{align}
	E_f = \frac{\delta}{3f}\left(2^{f+1} + 2^{2-f} - 6  \right) \label{maxerror}
\end{align}
and it takes its minimum value when $F_m=F_f$ as $E_f=0$. Since the choice of $F_m$ is a design parameter, one can use time sharing between the possible selections of $F_m$ to achieve any desired value for the expected number of false classifications between $E_f=0$ and $E_f$ in  \eqref{maxerror}.

\paragraph{Required number of tests:}
We first recall that, if we choose the sampling cluster formation level $F_m$, the required number of tests for selected individuals at that level for whom we design an $\mathcal{F}$-separable test matrix depends on the subtree that is composed of the first $m$ levels of the cluster formation tree $\mathcal{F}$. Note that the first $m$ levels of an exponentially split cluster formation tree is also an exponentially split cluster formation tree with $m$ levels. In \thref{theorem2} below, we focus on the sampling cluster formation choice at the bottom level, $F_m = F_f$ and characterize the \emph{exact} required number of tests to be between $f$ and $\frac{4}{3}f$. This implies that the required number of tests at level $F_f$ is $O(f)$, and thus, the required number of tests at level $F_m$ is $O(m)$. 

\begin{theorem} \thlabel{theorem2}
For an $f$ level exponentially split cluster formation tree, at level $f$, there exists an $\mathcal{F}$-separable test matrix, $\bm{X}$, with not more than $\frac{4}{3}f$ rows, i.e., an upper (achievable) bound for the number of required tests is $\frac{4}{3}(\log_2n + 1)$ for $n$ individuals. Conversely, this is also the capacity order-wise, since the number of required tests must be greater than $f$.
\end{theorem}

\begin{Proof}
By using the converse in \thref{lemma2}, we already know that the required number of tests is at least $f$ from \eqref{lemma2-eqn} since there are $\lambda_f = 2^f-1$ unique nodes and also $\lambda_{S_i^f} +1 = f$ for every subset $S_i^f$. This proves the converse part of the theorem.

In order to satisfy the sufficient conditions for the existence of an $\mathcal{F}$-separable matrix, each node in the tree must be represented by a $T$ length vector of sufficient Hamming weight, so that i) every descendant can be represented by a unique vector with positions of 1's being the subsets of the positions of 1's of their ancestor nodes, and ii) OR of vectors that are all descendants of a node must be equal to the vector of the ancestor node. In our proof, we show that, for exponentially split cluster formation trees, it is sufficient to check that we have sufficient number of rows in $\bm{X}$ to uniquely assign vectors to the bottom level nodes, i.e., the subsets $S_i^f$ at level $F_f$.

First, as we stated above, from the converse in \thref{lemma2}, an $\mathcal{F}$-separable test matrix of an exponentially split cluster formation tree with $f$ levels must have at least $f$ rows. However, for exponentially split cluster formation trees, this converse is not achievable: There are $2^{f-1}$ nodes at level $f$ but $\binom{f}{1}$ binary vectors with Hamming weight 1. Since for $f>3$, $\binom{f}{1}$ is less than $2^{f-1}$, we cannot assign distinct Hamming weight 1 vectors to the bottom level nodes. Thus, we need vectors with length longer than $f$. Now, assume that an achievable $\mathcal{F}$-separable test matrix has $f+k$ rows, where $k$ is a non-negative integer. Our objective in the remainder of the proof is to characterize this $k$ in terms of $f$.

We argue that if the number of nodes at the bottom level, which is equal to $2^{f-1}$, is less than $\sum_{i = 1}^{k+1}\binom{f+k}{i}$ then we can find an achievable $\mathcal{F}$-separable test matrix, i.e.,
\begin{align}
    \sum_{i = 1}^{k+1}\binom{f+k}{i} \geq 2^{f-1} \label{finalsuff}
\end{align}
is a sufficient condition for the existence of an achievable $\mathcal{F}$-separable test matrix for a given $(f,k)$ pair. Minimum $k$ that satisfies \eqref{finalsuff} will result in the minimum number of required tests $f+k$. In our construction, we assign each node at level $F_i$ a unique vector with Hamming weight $f+k+1-i$, except for the bottom level $F_f$. Since each node is assigned a unique vector, when moving from a level to one level down, descendant nodes must be assigned vectors that have Hamming weight at least 1 less than their ancestor node. At the bottom level, we use the remaining vectors with Hamming weight less than or equal to $k+1$. We choose minimum such $k$ for this construction, resulting in the minimum number of tests.

Before proving the achievability of this above construction, we first analyze the minimum $k$ that satisfies \eqref{finalsuff} in terms of $f$. We state and prove in \thref{applemma2} in Appendix that $k=f/3$ satisfies \eqref{finalsuff}, giving an upper bound for the minimum $k$, thus finalizing the first part of the achievability proof. This, in turn shows that, we can use all vectors of Hamming weight 1 through $k+1$ in the bottom level to represent all $2^{f-1}$ nodes at that level.

Next, we show that for the upper levels, our construction is achievable, i.e., we can find sufficiently many vectors of corresponding Hamming weights. By using \thref{lemma3} in  Appendix, and the fact that for $k \leq f/3$, when $f \geq 13$, we have
\begin{align} \label{finalsuff1}
	\binom{f+k}{k+2} \geq 2^{f-2}
\end{align}
which implies that, we can find unique vectors of Hamming weight $k+2$, to assign to the nodes at level $F_{f-1}$ (one level up from the bottom level). For the remaining levels below $\ceil{(f+k)/2}$, we have $\binom{f+k}{i} > \binom{f+k}{i+1}$ and the number of nodes decreases by half as we move upwards on the tree. Thus, we can find unique vectors to represent the nodes by increasing the Hamming weights by 1 at each level, which is the minimum increase of Hamming weights while moving upwards on the tree. For the remaining nodes, which are above the level $\ceil{(f+k)/2}$, we can use the lower bound for the binomial coefficient,
\begin{align}
	\binom{f+k}{i} \geq \left(\frac{f+k}{i}\right)^i \geq 2^i
\end{align}
to show that there are unique vectors of required weights at those levels as well.

Thus, there are sufficiently many unique vectors of appropriate Hamming weights at every level. Finally, we have to check whether or not there are sufficient number of unique vectors for every subtree of descendants of each node. In exponentially split cluster formation trees, due to the symmetry of the tree, any descendant subtrees of each node is again an exponentially split cluster formation tree. If we assume that $k$, where the number of rows of $\bm{X}$ is equal to $f+k$, satisfies \eqref{finalsuff} with $k$ being minimum such number, then every descendant subtree below the top level has parameters $(f-i,k)$ and we show in \thref{applemma2} in Appendix that they also satisfy the condition \eqref{finalsuff}. For $f$ values that are below the corresponding threshold in our proof steps (e.g., $f\geq 13$ threshold before (\ref{finalsuff1}) above), manual calculations yield the desired results. This proves the achievability part of the theorem.
\end{Proof}

\paragraph{Expected number of infections:} In an exponentially split cluster formation tree structure with $f$ levels, the expected total number of infections is,
\begin{align}
	\sum_{i = 1}^{f} \frac{1}{f} 2^{f-i}\delta = \frac{\delta}{f}(2^f-1) \label{expectedinf}
\end{align}
since $p_F(F_i) = 1/f$ and if $F=F_i$ then there are $2^{f-i}\delta$ infections. Thus, the expected number of infections is $O\left(\frac{n}{\log_2n}\right)$.

\paragraph{Comparison:} In order to compare our results for the exponentially split cluster formation trees with other results in the literature, for fairness, we focus on the zero-error case in our system model, which happens when $F_m=F_f$ is chosen. Resulting sampling function selects in total $2^{f-1}$ individuals and the resulting number of required tests is between $f$ and $\frac{4}{3}f$, i.e., $O(\log_2n)$, as proved in \thref{theorem2}. Note that, by performing at most $\frac{4}{3}f$ tests to $2^{f-1}$ individuals, we identify the infection status of $2^{f-1}\delta$ individuals with zero false classifications, which implies that the number of tests scales with the number of nodes at the bottom level, instead of the number of individuals in the system. This results in a gain scaled with $\delta$. However, in order to fairly compare our results with the results in the literature, we ignore this gain and compare the performance of the second step of our algorithm only, i.e., the identification of infection status of selected individuals only. To avoid confusion, let $\delta=1$, i.e., each cluster at the bottom level is an individual and thus, $n=2^{f-1}$.

From \eqref{expectedinf}, the expected number of infections in this system is $\frac{2^f-1}{f} = O(\frac{n}{\log_2n})$. When the infections scale faster than $\sqrt{n}$, as proved in \cite{RUSZINKO} (see also \cite{scarlettbook}), non-adaptive tests with zero-error criterion cannot perform better than individual testing. Since our algorithm results in $O(f) = O(\log_2n)$ tests, it outperforms all non-adaptive algorithms in the literature. Furthermore, we compare our results with Hwang's generalized binary splitting algorithm \cite{hwang_binary}, even though it is an adaptive algorithm and also it assumes the prior knowledge of exact number of infections. Hwang's algorithm results in a zero-error identification of $k$ infections among the population of $n$ individuals with $k\log _2(n/k)+O(k)$ tests and attains the capacity of adaptive group testing \cite{scarlettbook,hwang_binary,adaptivecapacity}. Since the number of infections takes $f$ values in the set $\{1, 2, 2^2,\dots,2^{f-1}\}$ uniformly randomly, the resulting mean value of the required number of tests when Hwang's generalized binary splitting algorithm is used is
\begin{align}
    \mathbb{E}[T_{\text{Hwang}}] &= \sum_{i=0}^{f-1}\frac{1}{f}\left(2^i\log_22^{f-1-i}\right) + O\left(\frac{n}{\log_2n}\right) \\
      &=\frac{f-1}{f}\sum_{i=0}^{f-1}2^i - \frac{1}{f}\sum_{i=0}^{f-1}i2^i + O\left(\frac{n}{\log_2n}\right)\\
      &=\frac{2^f-f-1}{f}+O\left(\frac{n}{\log_2n}\right)\\
      &=O\left(\frac{n}{\log_2n}\right)
\end{align} 

Thus, the expected number of tests when Hwang's generalized binary splitting algorithm is used scales as $O\left(\frac{n}{\log_2n}\right)$ which is much faster than our result of $O(\log_2n)$. We note that, Hwang's generalized binary splitting algorithm assumes the prior knowledge of exact number of infections, and is an adaptive algorithm, and further, we have ignored the gain of our algorithm in the first step (i.e., $\delta=1$). Despite these advantages given to it, our algorithm still outperforms Hwang's generalized binary splitting algorithm for exponentially split cluster formation trees.

\section{Numerical Results}
In this section, we present numerical results for the proposed two-step sampled group testing algorithm and compare our results with the existing results in the literature. In the first simulation environment, we focus on exponentially split cluster formation trees as presented in Section~\ref{sec5}, and in the second simulation environment, we consider an arbitrary random connection graph, as discussed in Section~\ref{sec2}, which does not satisfy the cluster formation tree assumption. In the first simulation environment, we verify our analytical results by focusing on exponentially split cluster formation trees. In the second simulation environment, we show that our ideas can be applied to arbitrary random connection graph based networks.

\subsection{Exponentially Split Cluster Formation Tree Based System} \label{sec61}

\begin{figure}[t]
    \centering
    \begin{subfigure}[t]{0.475\textwidth}
        \centering
        \epsfig{file=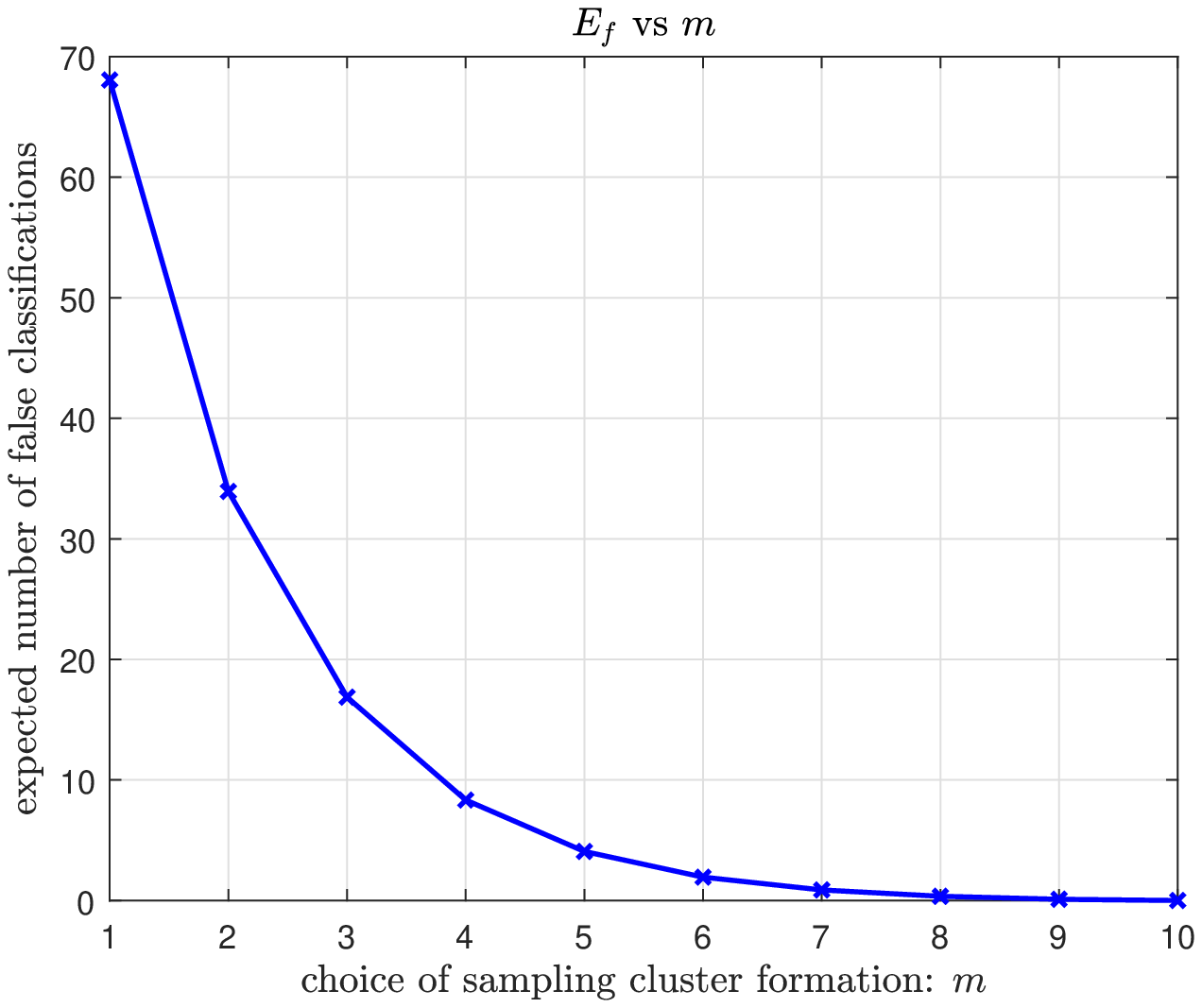,scale=0.5}
        \caption[]%
        {}
    \end{subfigure}%
    ~ 
    \begin{subfigure}[t]{0.475\textwidth}
        \centering
        \epsfig{file=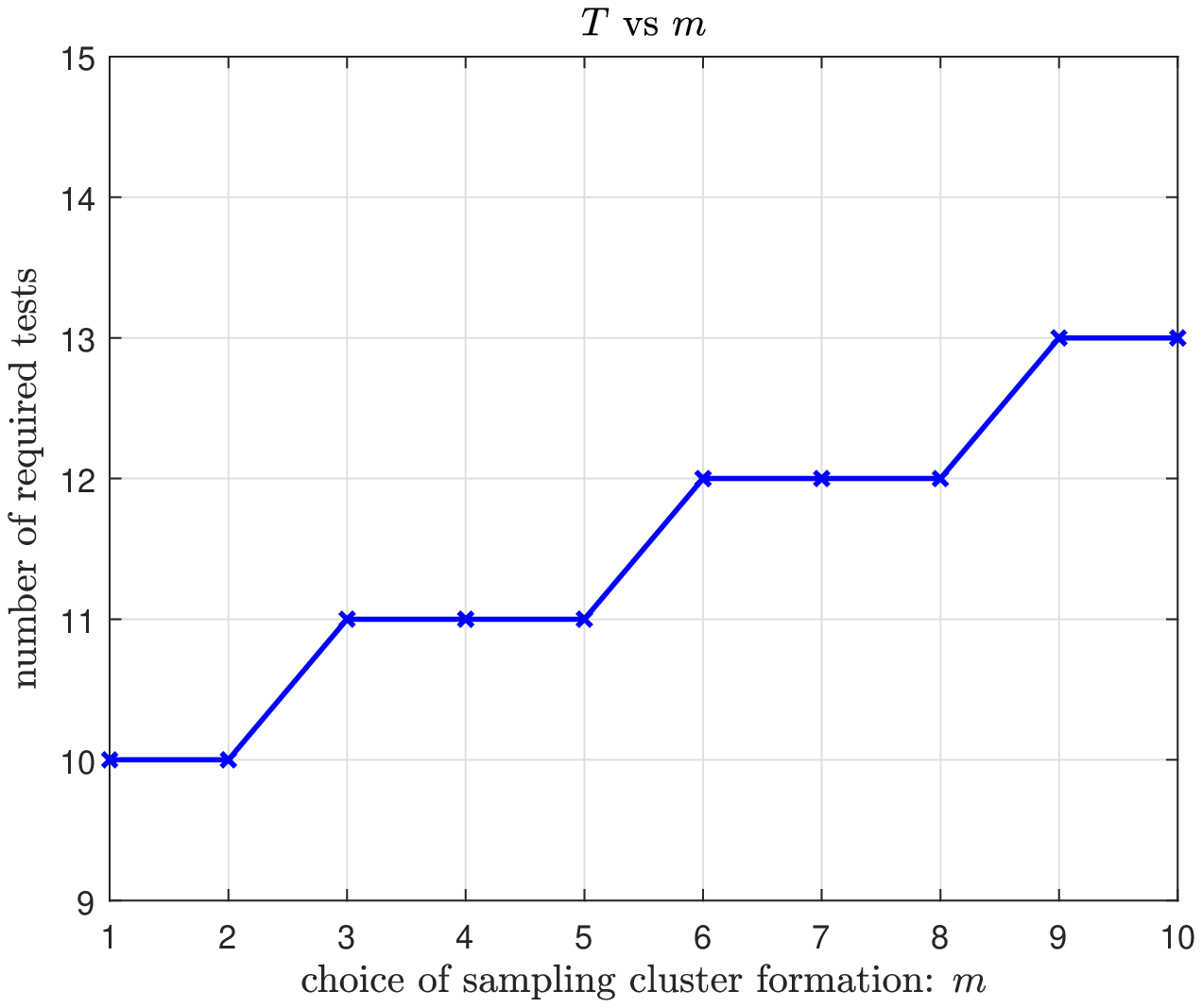,scale=0.5}
        \caption[]%
        {}
    \end{subfigure}
    \caption{(a) Expected number of false classifications vs the choice of sampling cluster formation $F_m$. (b) Required number of tests vs the choice of sampling cluster formation $F_m$.} 
    \label{num12}
\end{figure}

In the first simulation environment, we have an exponentially split cluster formation tree with $f=10$ levels and $\delta=1$ at the bottom level. For this system of $n=2^{f-1} \delta=512$ individuals, for each sampling cluster formation choice $F_m$, from $m=1$, i.e., the top level of the cluster formation tree, to $m=10$, i.e., the bottom level of the cluster formation tree, we calculate the expected number of false classifications and the minimum required number of tests. Note that the required number of tests is fixed for a fixed sampling cluster formation $F_m$, while the number of false classifications depends on the realization of the true cluster formation $F_\alpha$ and patient zero $Z$. In Figure~\ref{num12}(a), we plot the expected number of false classifications which meets the analytical expressions we found in Section~\ref{sec5}. While calculating the minimum number of required tests, for each choice of $F_m$, our program finds the minimum $T$ that satisfies the sufficient criteria that we presented in Section~\ref{sec4} and in the proof of \thref{theorem2}. We plot the minimum required number of tests in Figure~\ref{num12}(b). Note that, unlike the number of false classifications, for a fixed $F_m$, the number of required tests is fixed and thus, we do not repeat the simulations while calculating the required number of tests. The resulting non-adaptive test matrix $\bm{X}$ is fixed for a fixed $F_m$ and identifies the infection status of the individuals that are selected by $M$, with zero-error. 

Next for this network setting, we compare our zero-error construction results with the results of a variation of Hwang's generalized binary splitting algorithm \cite{adaptivecapacity,hwang_binary}, presented in \cite{allemann}, which further reduces the number of required tests by reducing the $O(k)$ term in the capacity expression of Hwang's algorithm. As we state in the comparison part of Section~\ref{sec5}, the required number of tests in our algorithm scales with $O(\log_2n)$, resulting in 13 tests at level $m=f=10$, as seen in Figure~\ref{num12}(b), while the average number of required tests for Hwang's algorithm scales as $O\left(\frac{n}{\log_2n}\right)$, and is approximately 172 in this case. Further, when we remove the assumption of known number of infections, we have to use the binary splitting algorithm presented originally in \cite{binarysplittingorig}, which results in a number of tests that is not lower than individual testing, i.e., $n=512$ tests in this case.

\begin{figure}[t]
    \begin{subfigure}{0.475\textwidth} 
        \centering
        \epsfig{file=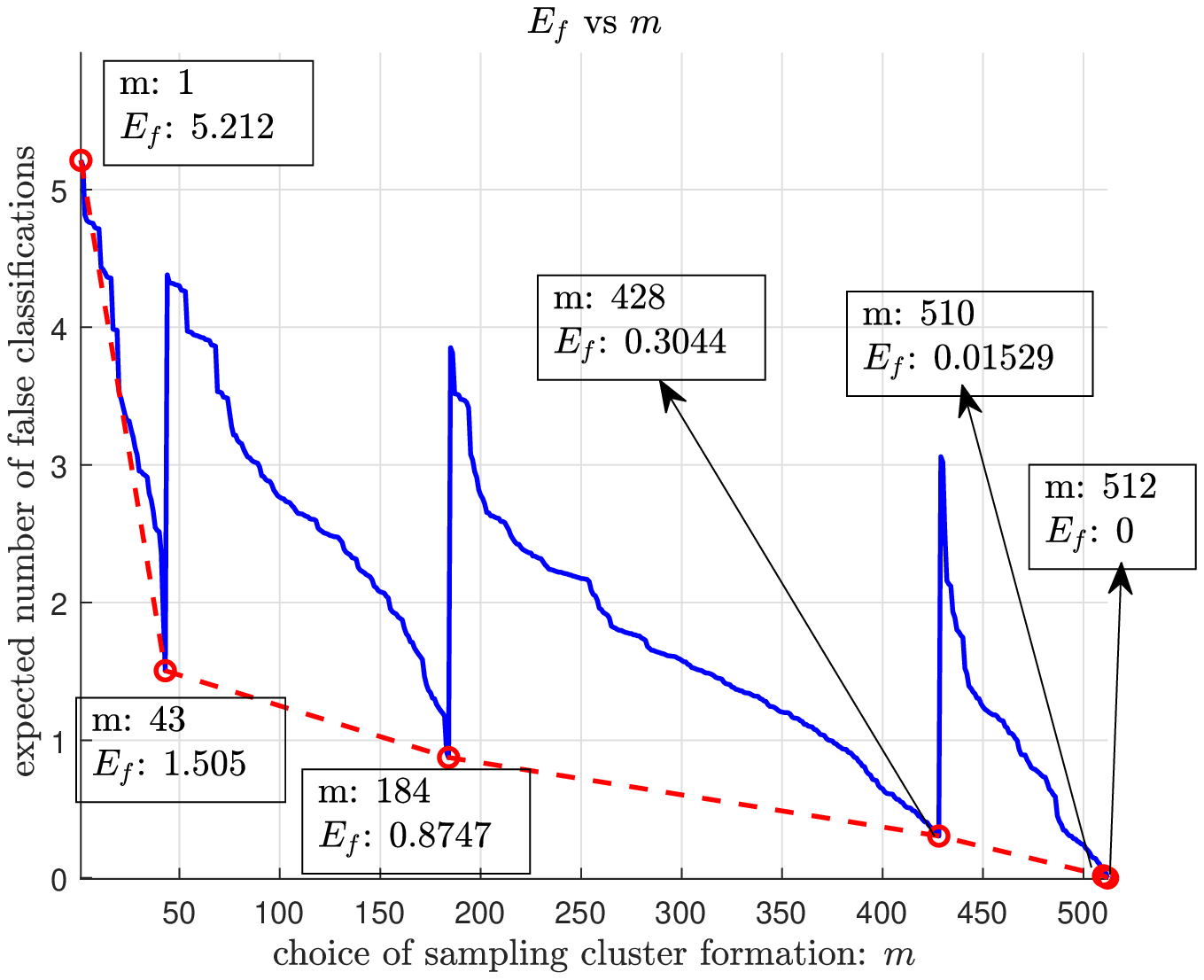,scale=0.5}
        \caption[]%
        {}
    \end{subfigure}
    \begin{subfigure}{0.475\textwidth}
        \centering
        \epsfig{file=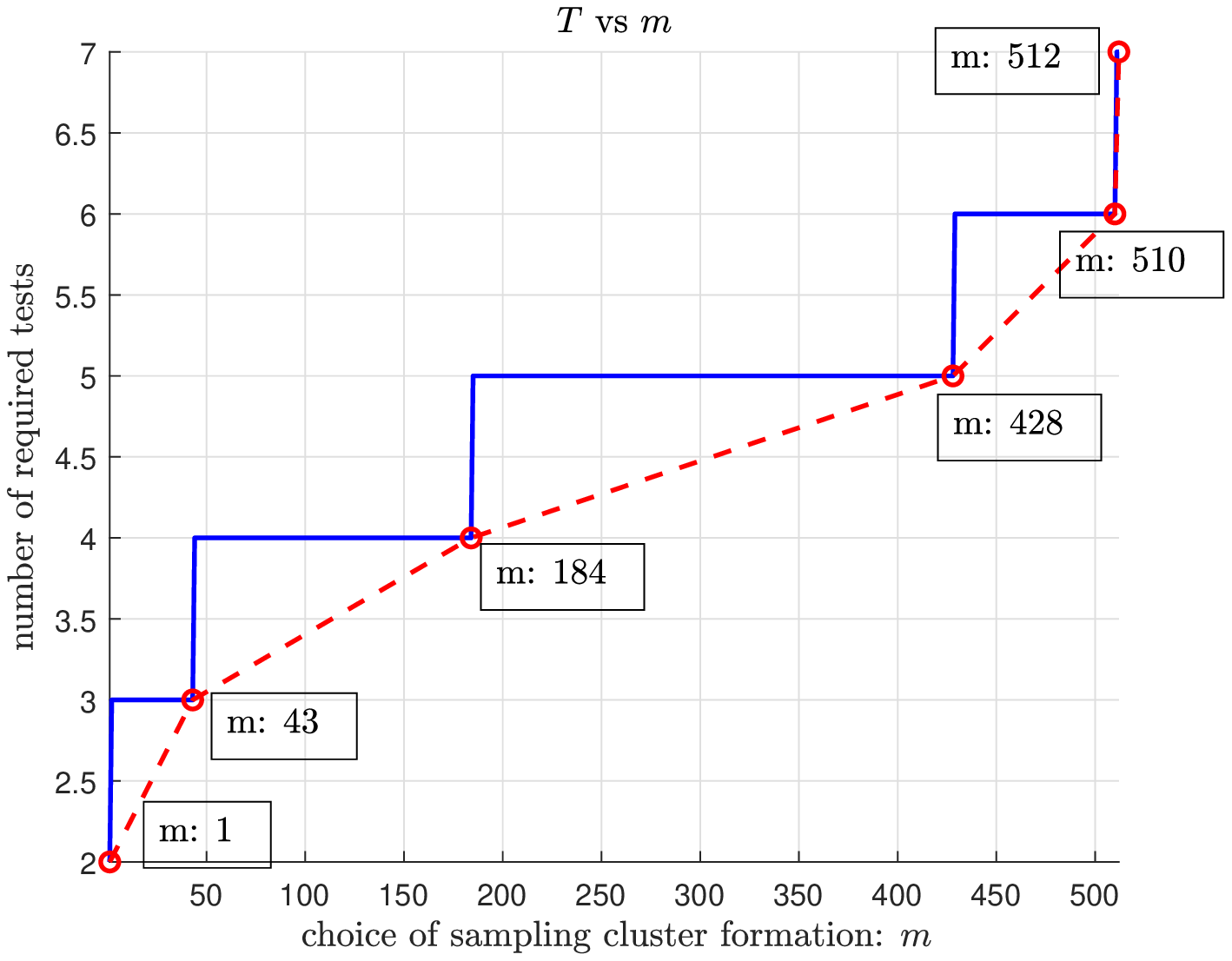,scale=0.5} 
        \caption[]%
        {}
    \end{subfigure}
    \begin{subfigure}{\textwidth}
        \centering
        \epsfig{file=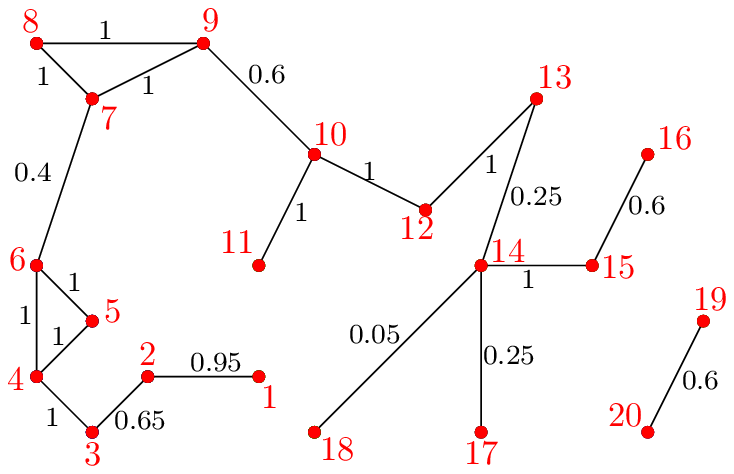,scale=1}
        \caption[]%
        {}
    \end{subfigure}
    \caption{(a) Expected number of false classifications vs the choice of sampling cluster formation $F_m$. (b) Required number of tests vs the choice of sampling cluster formation $F_m$. (c) Random connection graph.}
    \label{num345}
\end{figure}

\subsection{Arbitrary Random Connection Graph Based System}
In our second simulation environment, we present an arbitrary random connection graph $\mathscr{C}$ with 20 individuals, shown in Figure~\ref{num345}(c), where the edges realize independently with probabilities shown on them (zero probability edges are not shown). In this system, since each independent realization of 9 edges that can be either present or not results in a distinct cluster formation, in total, there are $2^9=512$ cluster formations that can be realized with positive probability. Note that this system with the random connection graph $\mathscr{C}$ does not yield a cluster formation tree, yet we still apply our ideas designed for cluster formation trees here. For each one of the 512 possible selections of $m$, we plot the corresponding expected number of false classifications in Figure~\ref{num345}(a) and the required number of tests in Figure~\ref{num345}(b) for our two-step sampled group testing algorithm.

In this simulation, for each possible choice of the sampling cluster formation $F_m$, we calculate the set of all possible infected sets $\mathcal{P}(K_M)$ for all possible choices of $M$ and calculate the resulting expected number of false classifications by also calculating $p_F$, the probability distribution of random cluster formations and select the optimal sampling function $M$. For the required number of tests, we find the minimum number of tests that satisfies the sufficient criteria that we presented in Section~\ref{sec4} in order to construct $\mathcal{F}$-separable matrices for this system. In our simulation environment, this procedure is done by brute force, since this system is not a cluster formation tree as in our system model and we cannot use the systematic results that we derived. This simulation demonstrates that the ideas presented can be generalized and applied to arbitrary random connection graph structures. 

Since the system here is arbitrary unlike the exponentially split cluster formation tree structure in the first simulation environment in Section~\ref{sec61}, the resulting expected number of false classifications is not monotonically decreasing when we sort the resulting required number of tests in the increasing order for the choices of $F_m$. In Figure~\ref{num345}(a), we mark the choices of sampling cluster formations that result in the minimum number of expected false classifications within each required number of test range. By using time sharing between these choices of the sampling cluster formations, dotted red lines between them can be achieved. The 6 corner points in Figure~\ref{num345}(a)-(b) correspond to the following cluster formations,
\begin{align}
    F_1   &= \{\{1\mbox{-}18\},\{19\mbox{-}20\}\}\\
    F_{43}  &= \{\{1\mbox{-}6\},\{7\mbox{-}13\},\{14\mbox{-}18\},\{19\mbox{-}20\}\} \label{f43}\\
    F_{184} &= \{\{1\mbox{-}6\},\{7\mbox{-}9\},\{10\mbox{-}13\},\{14\mbox{-}18\},\{19\},\{20\}\}\\
    F_{428} &= \{\{1\},\{2\},\{3\mbox{-}6\},\{7\mbox{-}9\},\{10\mbox{-}13\},\{14\mbox{-}17\},\{18\},\{19\},\{20\}\}\\
    F_{510} &= \{\{1,2\},\{3\mbox{-}6\},\{7\mbox{-}9\},\{10\mbox{-}13\},\{14,15\},\{16\},\{17\},\{18\},\{19\},\{20\}\}\\
    F_{512} &= \{\{1\},\{2\},\{3\mbox{-}6\},\{7\mbox{-}9\},\{10\mbox{-}13\},\{14,15\},\{16\},\{17\},\{18\},\{19\},\{20\}\}
\end{align}
For instance, $F_{43}$ in \eqref{f43} is composed of 4 clusters with $S_1^{43} = \{1,2,3,4,5,6\}$, $S_2^{43} = \{7,8,9,10,11,12,13\}$, $S_3^{43} = \{14,15,16,17,18\}$ and $S_4^{43} = \{19,20\}$. When $F_m=F_{43}$ is chosen as the sampling cluster formation, the resulting expected number of false classifications is $E_f=1.505$ and the required number of tests is $3$, as seen in Figure~\ref{num345}(a) and (b). For the sampling cluster formation choices which are not one of the six cluster formations listed above, these six cluster formations can be chosen to minimize the expected number of false classifications while keeping the required number of tests constant. For instance, all choices of $m$ between $m=2$ and $m=42$ result in required number of 3 tests as $m=43$ but yield a larger $E_f$ than what $m=43$ yields. 

For this system as well, we calculate the average number of required tests for Hwang's generalized binary splitting algorithm by using the results of \cite{adaptivecapacity, hwang_binary, allemann} as in the first simulation and find that the average number of required tests is 16.4 in this case. Similar to the first simulation environment, binary splitting algorithm presented originally in \cite{binarysplittingorig} which does not require the exact number of infections, cannot perform better than individual testing.

\section{Conclusions} 
In this paper, we introduced a novel infection spread model that consists of a random patient zero and a random connection graph, which corresponds to a non-identically distributed and correlated (non i.i.d.) infection status for individuals. We proposed a family of group testing algorithms, which we call \textit{two step sampled group testing algorithms}, and characterized their optimal parameters. We determined the optimal sampling function selection, derived expected false classifications, and proposed $\mathcal{F}$\textit{-separable non-adaptive group tests} which is a family of zero-error non-adaptive group testing algorithms that exploit a given random cluster formation structure. For a specific family of random cluster formations, which we call \textit{exponentially split cluster formation trees}, we calculated the expected number of false classifications and the required number of tests explicitly, by using our general results, and showed that our two-step sampled group testing algorithm outperforms all non-adaptive tests that do not exploit the cluster formation structure and Hwang's adaptive generalized binary splitting algorithm, even though our algorithm is non-adaptive and we ignore our gain from the first step of our two-step sampled group testing algorithm. Finally, our work has an important implication: in contrast to the prevalent belief about group testing that it is useful only when the infections are rare, our group testing algorithm shows that a considerable reduction in the number of required tests can be achieved by using the prior probabilistic knowledge about the connections between the individuals, even in scenarios with significantly high number of infections.

\section{Appendix}
\begin{lemma} \thlabel{applemma2}
    Minimum $k$ that satisfies
    \begin{align}
    	\sum_{i = 1}^{k+1}\binom{f+k}{i} \geq 2^{f-1} \label{lem2statement}
    \end{align}
    is upper bounded by $f/3$.
\end{lemma}
\begin{Proof}
    We prove the statement of the lemma by showing that the pair $(f,k)=(f,f/3)$ satisfies \eqref{lem2statement}. We first consider the left hand side of (\ref{lem2statement}) when $f$ is incremented by 1 for fixed $k$, and write it as
    \begin{align}
    	\sum_{i = 1}^{k+1}\binom{f+k+1}{i} &= 2\sum_{i = 1}^{k+1}\binom{f+k}{i} + 1 - \binom{f+k}{k+1} \label{incrementf}
    \end{align}
    which follows by using the identity $\binom{a}{b}=\binom{a-1}{b-1}+\binom{a-1}{b}$.
    
   Second, we prove the following statement for $k \geq 1$,
    \begin{align}
        \sum_{i=1}^{k+1}\binom{4k}{i} \geq 2^{3k-1} \label{lem2ind}
    \end{align}
    Note that, when $k=f/3$, \eqref{lem2ind} is equivalent to \eqref{lem2statement} for $f$ values that are divisible by 3. For $f$ values that are not divisible by 3, since the pairs $(f-1,k)$ and $(f-2,k)$ satisfy \eqref{lem2statement} when the pair $(f,k)$ satisfies \eqref{lem2statement}, by \eqref{incrementf}, it suffices to prove the statement in \eqref{lem2ind}.

    We prove \eqref{lem2ind} by induction on $k$. For $k=1$, the inequality holds. Assume that the inequality holds for a $k \geq 1$, then we show that it also holds for $k+1$. In the lines below, we use the identity $\binom{a}{b}=\binom{a-1}{b-1}+\binom{a-1}{b}$ recursively,
    \begin{align}
        \sum_{i=1}^{k+2}\binom{4k+4}{i} &= \sum_{i=1}^{k+2}\binom{4k+3}{i} + \sum_{i=1}^{k+2}\binom{4k+3}{i-1} \\
        &= \sum_{i=1}^{k+2}\binom{4k+2}{i} + \sum_{i=1}^{k+2}\binom{4k+2}{i-1} + 1 + \sum_{i=1}^{k+1}\binom{4k+2}{i} +\sum_{i=1}^{k+1}\binom{4k+2}{i-1}\\
        &\quad\vdots \nonumber\\
        &=9\sum_{i=1}^{k+1}\binom{4k}{i}-5\binom{4k}{k+1}+\binom{4k}{k+2}+4\binom{4k}{k-1} + 5\binom{4k}{k-2}+ A \label{recursionend}\\
        &=9\sum_{i=1}^{k+1}\binom{4k}{i} - \frac{2k+11}{k+2}\binom{4k}{k+1}+4\binom{4k}{k-1} + 5\binom{4k}{k-2}+ A\\
        &=8\sum_{i=1}^{k+1}\binom{4k}{i}- \frac{k+9}{k+2}\binom{4k}{k+1}+\binom{4k}{k}+5\binom{4k}{k-1} + 6\binom{4k}{k-2}+ A'\\
        &=8\sum_{i=1}^{k+1}\binom{4k}{i}+3\binom{4k}{k-2}+ A''\\
        &\geq 2^{3k+2} \label{lemma2final}
    \end{align}
    where $A, A', A''$ are positive terms that are $o\left(\binom{4k}{k-2}\right)$, and we use the identity $\binom{a}{b}=\frac{a-b+1}{b}\binom{a}{b-1}$ after equation \eqref{recursionend} to eliminate the negative $\binom{4k}{k+1}$ term. Inequality \eqref{lemma2final} follows from the induction assumption. This proves the statement for $k+1$ and completes the proof.
\end{Proof}

\begin{lemma} \thlabel{lemma3}
	When $k \leq \frac{2n-8}{5}$, the following inequality holds
	\begin{align}
		\frac{1}{2}\sum_{i = 1}^{k} \binom{n}{i} < \binom{n}{k+1}
	\end{align}
\end{lemma}
\begin{Proof}
	We prove the lemma by induction over $k$. First, note that the inequality holds when $k=1$,
	\begin{align}
		\frac{1}{2} \binom{n}{1} < \binom{n}{2}
	\end{align}
	Then, assume that the statement is true for $k$. Now we check the statement for $k+1$,
	\begin{align}
		\frac{1}{2}\sum_{i = 1}^{k+1} \binom{n}{i} &< \frac{3}{2}\binom{n}{k+1}\label{l31}\\ 
		&\leq \frac{n-k-1}{k+2} \binom{n}{k+1}\label{l32} \\ 
		&= \binom{n}{k+2}
	\end{align}
	where \eqref{l31} follows from the induction assumption and \eqref{l32} is because $k \leq \frac{2n-8}{5}$. This proves the statement for $k+1$ and completes the proof.
\end{Proof}

\bibliographystyle{unsrt}
\bibliography{references_grouptesting}

\end{document}